\documentclass{article}
\usepackage{amsmath,amssymb}
\usepackage[dvips,dvipdfmx]{graphicx,color}
\usepackage{here}
\title{Spatial pattern of discrete and ultradiscrete Gray-Scott model}
\author{Keisuke Matsuya$^1$ and Mikio Murata$^2$\vspace{4pt}\\${}^1$Graduate School of Mathematical Sciences,\\\vspace{2pt}University of Tokyo, 3-8-1 Komaba, Tokyo 153-8914, Japan\vspace{4pt}\\${}^2$Division of Mathematical Science Institute of Symbiotic, \\\vspace{2pt}Science and Technology,\\\vspace{2pt}Tokyo University of Agriculture and Technology, \\\vspace{2pt}3-5-8 Naka Town, Koganei City, Tokyo 184-8588, Japan}
\date{}
\begin{document}
\maketitle
\section{Introduction}
\label{s1}
Discretization is a procedure to get difference equations with some parameters from given differential equations.
The difference equations change to the differential equations with limits of the parameters.
This procedure is often used when one computes differential equations numerically.

Ultradiscretization \cite{4} is a limiting procedure transforming given difference equations into other difference equations which consist of addition, subtraction and maximum including cellular automata.
In this procedure, a dependent variable $u_n$ in a given equation is replaced by
\begin{equation}\label{ud}
u_n=\exp{\left(\frac{U_n}{\varepsilon}\right)},
\end{equation}
where $\varepsilon$ is a positive parameter.
Then, we apply $\varepsilon\log$ to both sides of \eqref{ud} and take the limit $\varepsilon\to+0$.
Using identity
\begin{equation*}
\displaystyle{\lim_{\varepsilon\to+0}{\varepsilon\log{(e^{U/\varepsilon}+e^{V/\varepsilon})}}=\max{(U,V)}}
\end{equation*}
and exponential laws, we find that multiplication, division and addition for the original variables are replaced by addition, subtraction and maximum for the new ones, respectively.
In this way, the original difference equation is approximated by a piecewise linear equation.

Properties of solutions for the difference and the ultradiscrete equations might be different from those of solutions for the differential equations.
If discretization and ultradiscretization are pursued nicely, some properties of solutions for differential equations are preserved.
Indeed, such difference and ultradiscrete equations obtained from differential equations are studied in the case of integrable equations such as soliton equations.
They preserve the essential properties of the original soliton equations, such as the structure of exact solutions \cite{5,6}.
There are few cases that discretization and ultradiscretization whose solutions inherit similar properties of solutions for differential equations which are not integrable.
The aim of this study is to get discretization and ultradiscretization of non-integrable equations and to elucidate correspondences between differential equations, difference equations and ultradiscrete equations.
In this paper, a systematic procedure to construct systems of difference equations and cellular automaton models is given.

Gray-Scott model \cite{1} is a variant of the autocatalytic model.
Basically it considers the reactions
\begin{equation*}
\begin{cases}
\text{U}+2\text{V}\to3\text{V},\\
\text{V}\to\text{P},
\end{cases}
\end{equation*}
in an open flow reactor where U is continuously supplied, and the product P removed.

A mathematical model of the reactions bellow is the following system of partial differential equations:
\begin{equation}\label{gs}
\begin{cases}
\displaystyle{\frac{\partial u}{\partial t} = \Delta u - uv^2 + a(1-u)},\vspace{4pt}\\
\displaystyle{\frac{\partial v}{\partial t} = D_v \Delta v + uv^2 - bv},
\end{cases}
\end{equation}
where $u:=u(t,\vec{x}),\ v:=v(t,\vec{x}),\ t\ge0,\ \vec{x}\in\mathbb{R}^d$ and $D_v,\ a$ and $b$ are positive constants.
$\Delta$ is $d$-dimensional Laplacian.
The solutions of this system represent spatial patterns.
Changing not only an initial condition but also parameters, various patterns are observed\cite{9,7,8}.

Considering \eqref{gs} with a spatially uniform initial condition, we get
\begin{equation}\label{ode}
\begin{cases}
\displaystyle{\frac{du}{dt}= -uv^2 + a(1-u)},\vspace{4pt}\\
\displaystyle{\frac{dv}{dt}= uv^2 - bv}.
\end{cases}
\end{equation}
Solving simultaneous equations, we get equilibrium points of \eqref{ode} as follow:
\begin{eqnarray*}
P_{c,0}:(u_{c,0},v_{c,0})&=&(1,0),\\
P_{c,\pm}:(u_{c,\pm},v_{c,\pm})&=&\left(\frac{1}{2}\left(1\mp\sqrt{1-\frac{4b^2}{a}}\right),\frac{a}{2b}\left(1\pm\sqrt{1-\frac{4b^2}{a}}\right)\right).
\end{eqnarray*}
$P_{c,+}$ and $P_{c,-}$ emerge when $a-4b^2\ge0$ is held.
$P_{c,0}$ is asymptotically stable.
$P_{c,-}$ is unstable.
$P_{c,+}$ is asymptotically stable if the following inequality
\begin{equation}\label{ineq}
b-\frac{a^2}{2b^2}\left(1+\sqrt{1-\frac{4b^2}{a}}\right)<0
\end{equation}
is held.

In numerical computation of \eqref{gs}, we have to discretize it and consider a system of partial difference equations.
A naive discretization would be to replace $t$-differentials with forward differences and Laplacians with central differences such that \eqref{gs} turns into
\begin{equation}\label{dgs0}
\begin{cases}
\displaystyle{\frac{u_{n+1}^{\vec{j}}-u_n^{\vec{j}}}{\delta}=\sum^d_{k=1}{\frac{u_n^{\vec{j}+\vec{e}_k}-2u_n^{\vec{j}}+u_n^{\vec{j}-\vec{e}_k}}{\xi^2}}-u_n^{\vec{j}}(v_n^{\vec{j}})^2 + a(1-u_n^{\vec{j}})},\\
\displaystyle{\frac{v_{n+1}^{\vec{j}}-v_n^{\vec{j}}}{\delta}=D_v\sum^d_{k=1}{\frac{v_n^{\vec{j}+\vec{e}_k}-2v_n^{\vec{j}}+v_n^{\vec{j}-\vec{e}_k}}{\xi^2}}+u_n^{\vec{j}}(v_n^{\vec{j}})^2 - bv_n^{\vec{j}}},
\end{cases}
\end{equation}
where $u(n,\vec{j})(=:u_n^{\vec{j}}),\ v(n,\vec{j})(=:v_n^{\vec{j}}):\mathbb{Z}_{\ge0}\times\mathbb{Z}^d\to\mathbb{R}$, for positive constants $\delta$ and $\xi$, and where $\vec{e}_k\in\mathbb{Z}^d$ is a unit vector whose $k$th component is 1 and whose other components are 0.

Considering \eqref{dgs0} with a spatial uniform initial condition, we get a system of difference equations.
We get similar equilibrium points of \eqref{ode} but the stability of the equilibrium point $(1,0)$ is different.
If the parameter $\delta$ is sufficiently large, $(1,0)$ is unstable.
This case is different from the case of \eqref{gs}.

Since there are subtractions in \eqref{dgs0}, we cannot ultradiscretize \eqref{dgs0}.
Indeed, following limit
\begin{equation*}
\lim\limits_{\varepsilon\to+0}{\varepsilon\log(e^{U/\varepsilon}-e^{V/\varepsilon})}
\end{equation*}
does not always exist.
We can transform \eqref{dgs0} without subtractions and ultradiscretize the equations, but the obtained equations are not evolution equations.
This situation is inconvenient to investigate if solutions represent spatial patterns.

In this article, we propose discretization of \eqref{gs} which can be ultradiscretized and investigate solutions of discretization and ultradiscretization of \eqref{gs}.
Solutions of the discretization and ultradiscretization give various patterns by changing the parameters in the equations.

In Section \ref{s2}, we present a system of partial difference equations whose continuous limit equals \eqref{gs}, and consider the solutions of the discretization.
In Section \ref{s3}, we present the ultradiscretization of the system of partial difference equations treated in Section 2 and consider the solutions of the ultradiscretization.
Finally concluding remarks are given in Section \ref{s4}.
\section{Discrete Gray-Scott model}
\label{s2}
In this section, we discretize \eqref{gs} and investigate solutions.
\subsection{Discretization of Gray-Scott model}
\label{ss2.1}
Since it is more convenient to consider the ultradiscretization, we take the scaling $w:=v+1$ which changes \eqref{gs} to
\begin{equation}\label{gs2}
\begin{cases}
\displaystyle{\frac{\partial u}{\partial t} = \Delta u - u(w-1)^2 + a(1-u)},\vspace{4pt}\\
\displaystyle{\frac{\partial w}{\partial t} = D_v\Delta w + u(w-1)^2 -b(w-1)}.
\end{cases}
\end{equation}
First we consider the discretization of following system of ordinary differential equations:
\begin{equation}\label{ode2}
\begin{cases}
\displaystyle{\frac{du}{dt} = -u(w-1)^2 + a(1-u)},\vspace{4pt}\\
\displaystyle{\frac{dw}{dt} = u(w-1)^2 - b(w-1)}.
\end{cases}
\end{equation}
We consider the following system of difference equations:
\begin{equation}\label{dode}
\begin{cases}
\displaystyle{u_{n+1} = \frac{u_n+\delta(2u_nw_{n+1}+a)}{1+\delta\{(w_{n+1})^2+a+1\}}},\vspace{2pt}\\
\displaystyle{w_{n+1}=\frac{w_n+\delta[u_n\{(w_n)^2+1\}+b]}{1+\delta(2u_n+b)}},
\end{cases}
\end{equation}
where $n\in\mathbb{Z}_{\ge0},\ \delta>0.$
The method of discretization is same to that used in \cite{2,3}

If there exists smooth functions $u(t),\ w(t)\ (t\ge0)$ that satisfy $u(\delta n)=u_n,\ w(\delta n)=w_n$, we find
\begin{equation*}
\begin{cases}
\displaystyle{\frac{u(t+\delta)-u(t)}{\delta}=-u(t)w(t+\delta)^2+a(1-u(t))+o(\delta)},\vspace{4pt}\\
\displaystyle{\frac{w(t+\delta)-w(t)}{\delta}=u(t)w(t)^2-b(w(t)-1)+o(\delta)}.
\end{cases}
\end{equation*}
Taking the limit $\delta\to+0$, we obtain the system of differential equations \eqref{ode2}.
Thus, \eqref{dode} can be regarded as a discretization of \eqref{ode2}.
Using \eqref{ode2}, we can construct a system of partial difference equations:
\begin{equation}\label{dgs}
\begin{cases}
\displaystyle{u_{n+1}^{\vec{j}} = \frac{m_p(u_n^{\vec{j}})+\delta(2m_p(u_n^{\vec{j}})w_{n+1}^{\vec{j}}+a)}{1+\delta\{(w_{n+1}^{\vec{j}})^2+a+1\}}},\vspace{2pt}\\
\displaystyle{w_{n+1}^{\vec{j}}=\frac{m_q(w_n^{\vec{j}})+\delta\{m_p(u_n^{\vec{j}})(m_q(w_n^{\vec{j}})^2+1)+b\}}{1+\delta(2m_p(u_n^{\vec{j}})+b)}},
\end{cases}
\end{equation}
where $n\in\mathbb{Z}_{\ge0},\ \vec{j}\in\mathbb{Z}^d$ and
\begin{equation*}
m_p(f^{\vec{j}}):=\sum^d_{k=1}{\frac{f^{\vec{j}+p\vec{e}_k}+f^{\vec{j}-p\vec{e}_k}}{2d}}\ (p\in\mathbb{Z}_{>0}).
\end{equation*}
Since \eqref{dgs} is equivalent to
\begin{equation*}
\begin{cases}
\displaystyle{\frac{u_{n+1}^{\vec{j}}-u_n^{\vec{j}}}{\delta}=\sum^d_{k=1}\frac{u_n^{\vec{j}+p\vec{e}_k}-2u_n^{\vec{j}}+u_n^{\vec{j}-p\vec{e}_k}}{(p\xi)^2}-m_p(u_n^{\vec{j}})(w_{n+1}^{\vec{j}})^2}\\
~~~~~~~~~~~~~~~~~~~~~~~~~~~~~~~~~~~~~~~~~~~~~~~~~~~~~~~~+a(1-m_p(u_n^{\vec{j}}))+o(\delta),\\
\displaystyle{\frac{w_{n+1}^{\vec{j}}-w_n^{\vec{j}}}{\delta}=\frac{q^2}{p^2}\sum^d_{k=1}\frac{w_n^{\vec{j}+q\vec{e}_k}-2w_n^{\vec{j}}+w_n^{\vec{j}-q\vec{e}_k}}{(q\xi)^2}+m_p(u_n^{\vec{j}})m_q(w_n^{\vec{j}})^2}\\
~~~~~~~~~~~~~~~~~~~~~~~~~~~~~~~~~~~~~~~~~~~~~~~~~~~~~~~~-b(m_q(w_n^{\vec{j}})-1)+o(\delta),
\end{cases}
\end{equation*}
where $\xi:=\sqrt{2d\delta}/p$, if there exists smooth functions $u(t,\vec{x}),\ w(t,\vec{x})\ (t\ge0,\ \vec{x}\in\mathbb{R}^d)$ that satisfy $u(\delta n,\xi\vec{j})=u_n^{\vec{j}}$ and $w(\delta n,\xi\vec{j})=w_n^{\vec{j}}$, we obtain \eqref{gs2} where $D_v=(q/p)^2$ with the limit $\delta\to0$.
Solving simultaneous equations, we get equilibrium points of \eqref{dode}:
\begin{eqnarray*}
P_{d,0}:(u_{d,0},w_{d,0})&=&(1,1),\\
P_{d,\pm}:(u_{d,\pm},w_{d,\pm})&=& \left(\frac{1}{2}\left(1\mp\sqrt{1-\frac{4b^2}{a}}\right),1+\frac{a}{2b}\left(1\pm\sqrt{1-\frac{4b^2}{a}}\right)\right).
\end{eqnarray*}
$P_{d,0}$ is asymptotically stable.
$P_{d,-}$ is unstable.
$P_{d,+}$ is asymptotically stable if the following inequality
\begin{equation*}
b-\frac{a^2}{2b^2}\left(1+\sqrt{1-\frac{4b^2}{a}}\right)+\delta\left[(2b-a)\left\{1+\frac{a}{2b}\left(1-\sqrt{1+\frac{4b^2}{a}}\right)\right\}-a\right]<0
\end{equation*}
is held.
The coefficient of $\delta^0$ in left hand side is same to the left hand side of \eqref{ineq}.
These equilibrium points are same as those of continuous case \eqref{ode} regardless of the value of $\delta$.

\subsection{Solutions for the discrete Gray-Scott model}
\label{ss2.2}
Now, let $d=1,\ p=3,\ q=1,\ \delta=0.1$ and
\begin{eqnarray*}
u_0^j=
\begin{cases}
1-0.3 \cos\left(\displaystyle{\frac{\pi j}{50}}\right)\ |j|\le25,\\
1~~~~~~~~~~~~~~~~~~~~~~|j|>25,
\end{cases}
\\
w_0^j=
\begin{cases}
1+0.5 \cos\left(\displaystyle{\frac{\pi j}{50}}\right)\ |j|\le25,\\
1~~~~~~~~~~~~~~~~~~~~~~|j|>25.
\end{cases}
\end{eqnarray*}
If one plots the solutions of \eqref{dgs} with a periodic boundary condition, following patterns are observed.
The horizontal axis is for space variable $j$.
The vertical axis is for time variable $n$.
The height means the value of $w^j_n$.
\begin{figure}[H]
\begin{minipage}{0.48\hsize}
\begin{center}
\includegraphics[scale=0.3]{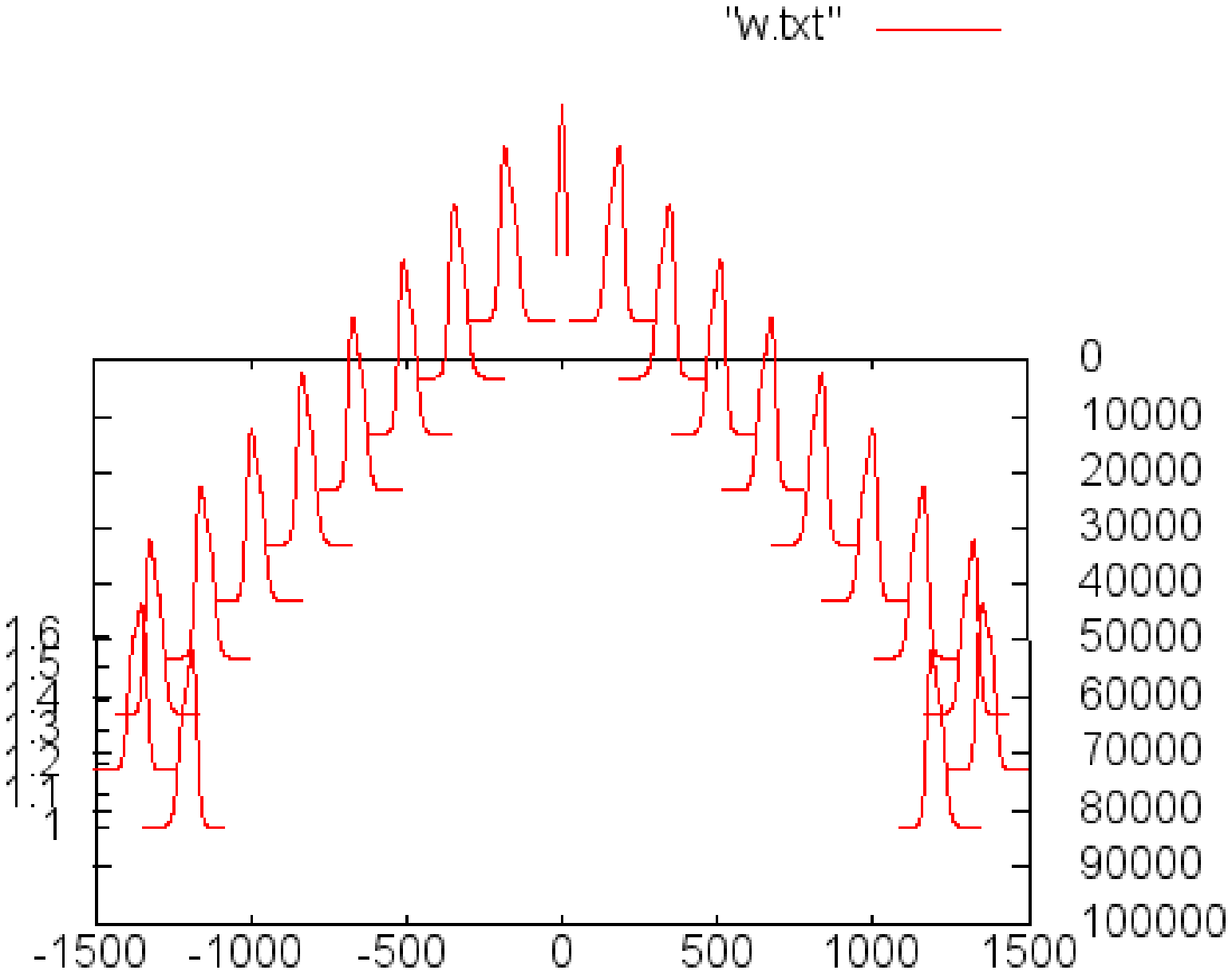}
\end{center}
\caption{$a=0.03,\ b=0.10$}
\label{1}
\end{minipage}
\begin{minipage}{0.48\hsize}
\begin{center}
\includegraphics[scale=0.3]{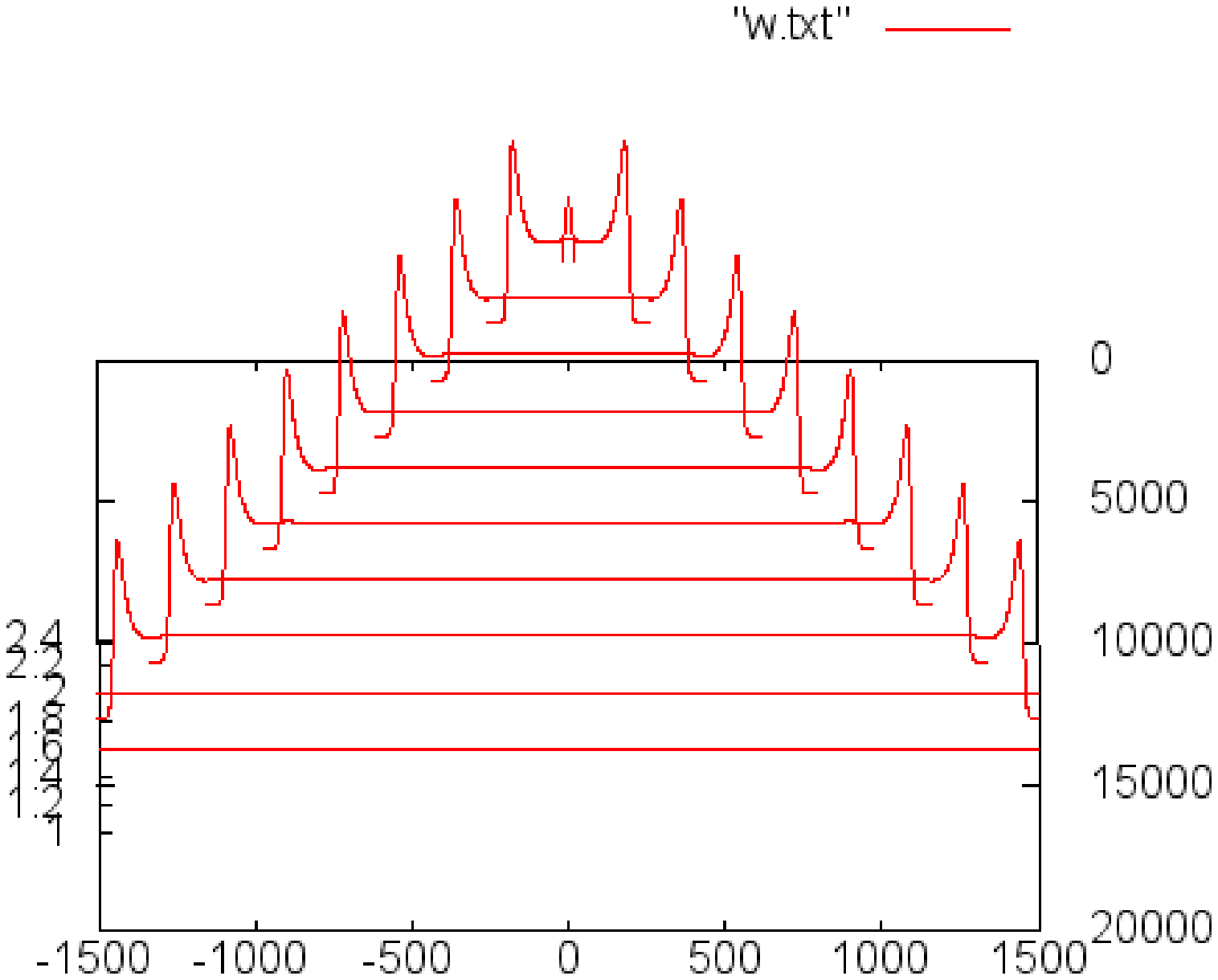}
\end{center}
\caption{$a=0.04,\ b=0.06$}
\label{2}
\end{minipage}
\end{figure}
In Figure \ref{1}, a peak split into two peaks and two peaks move opposite side.
We took a periodic boundary condition so that it is observed that two peaks pass each other.
In Figure \ref{2}, a similar situation of Figure \ref{1} is observed.
Between two peaks, values of $(u,w)$ converge to the stable equilibrium point $P_{d,-}$.
Moreover, two peaks vanish, when they collide.
\begin{figure}[H]
\begin{minipage}{0.48\hsize}
\begin{center}
\includegraphics[scale=0.3]{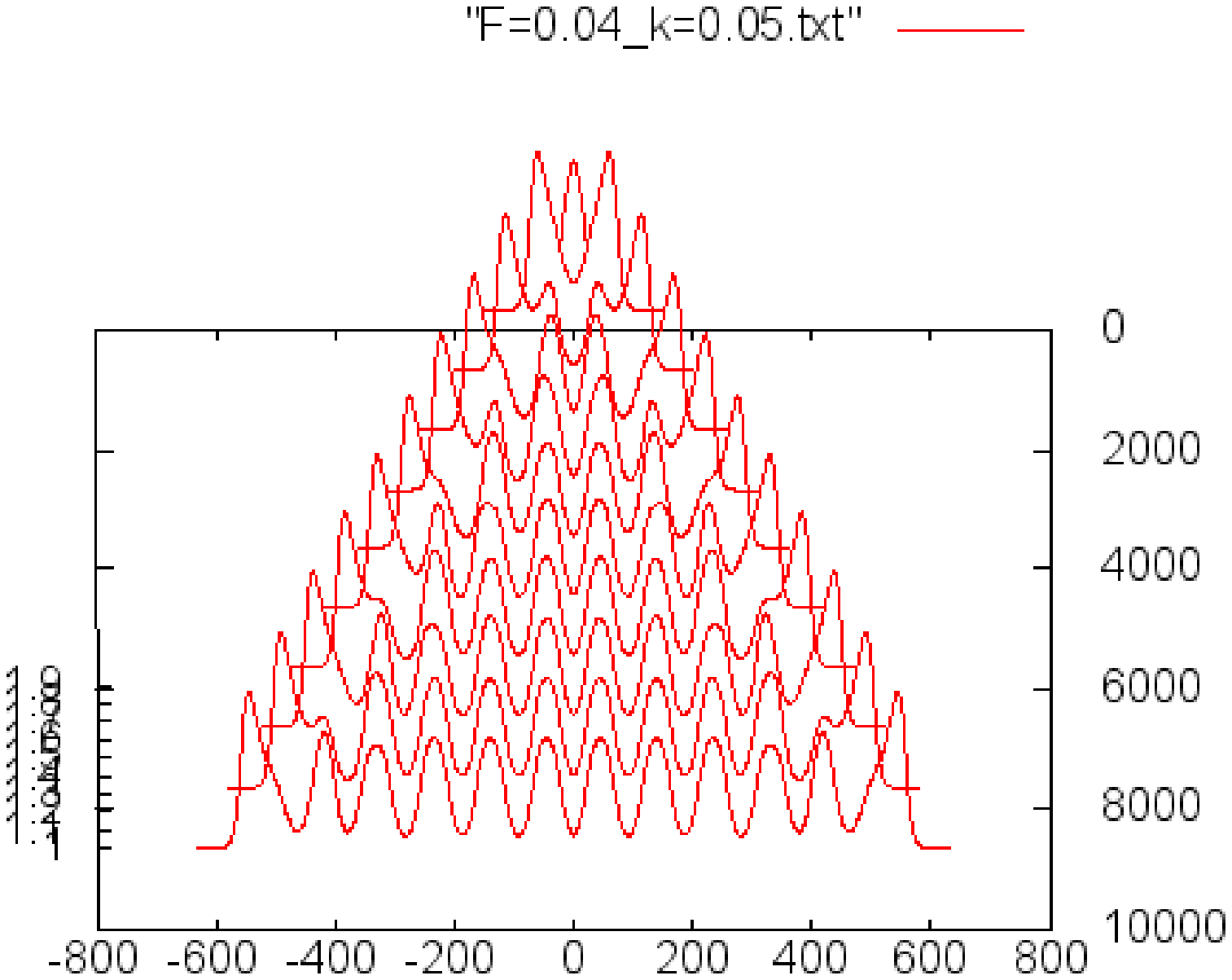}
\end{center}
\caption{$a=0.04,\ b=0.09$}
\label{3}
\end{minipage}
\begin{minipage}{0.48\hsize}
\begin{center}
\includegraphics[scale=0.3]{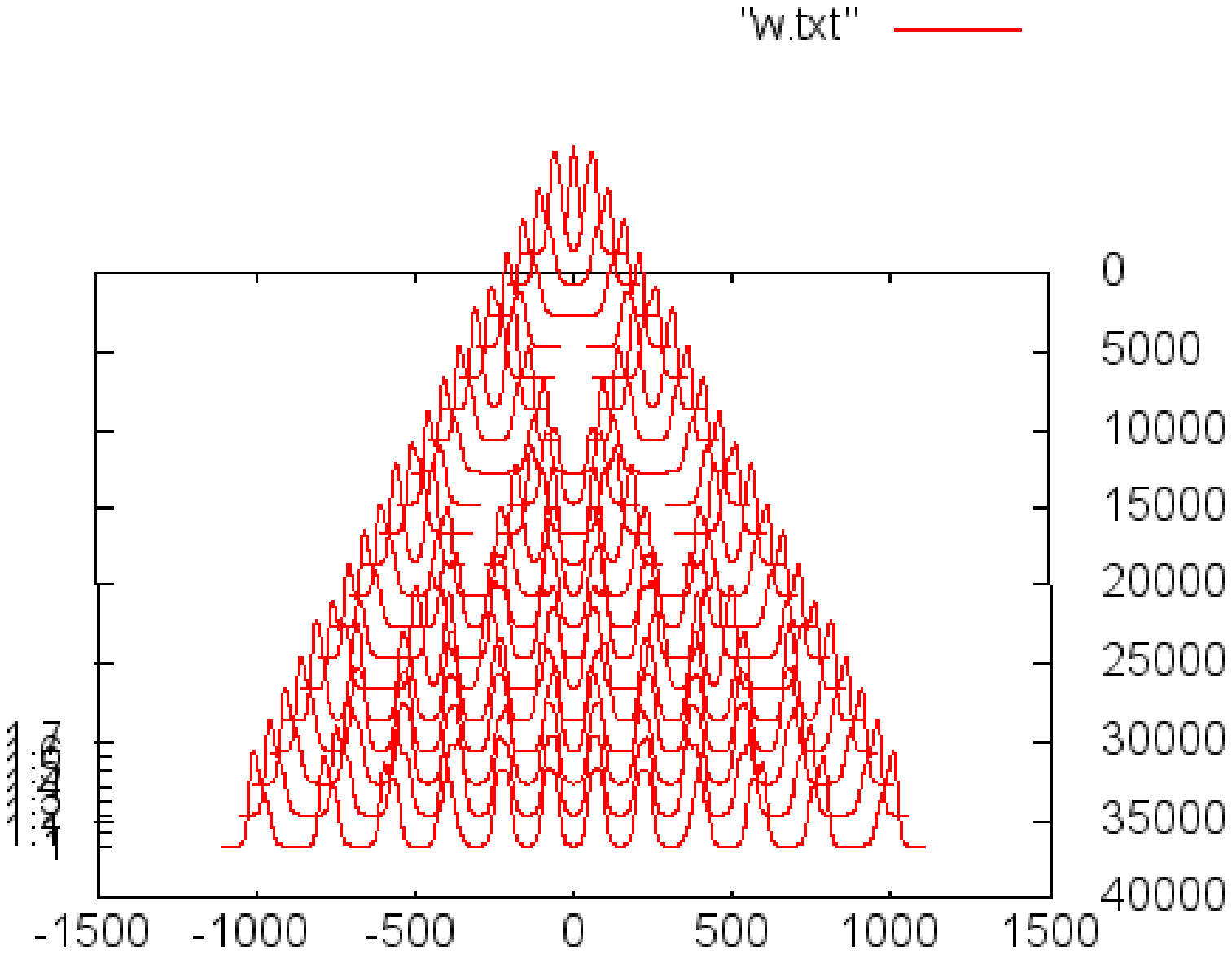}
\end{center}
\caption{$a=0.04,\ b=0.11$}
\label{4}
\end{minipage}
\end{figure}
In Figure \ref{3} and \ref{4}, a peak split into a two peaks several times and a self-replicating pattern is observed.
\begin{figure}[H]
\begin{minipage}{0.48\hsize}
\begin{center}
\includegraphics[scale=0.3]{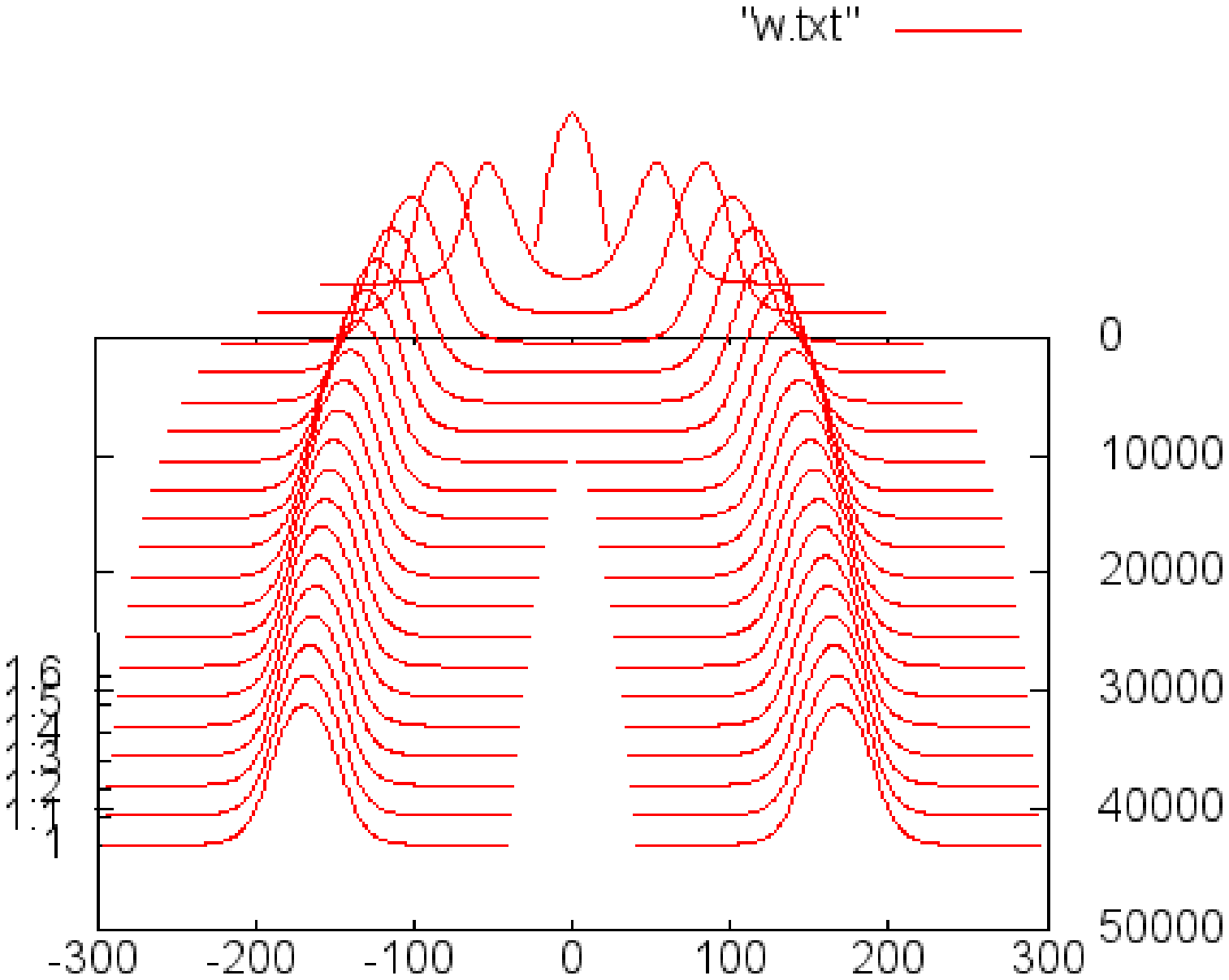}
\end{center}
\caption{$a=0.02,\ b=0.09$}
\label{5}
\end{minipage}
\begin{minipage}{0.48\hsize}
\begin{center}
\includegraphics[scale=0.3]{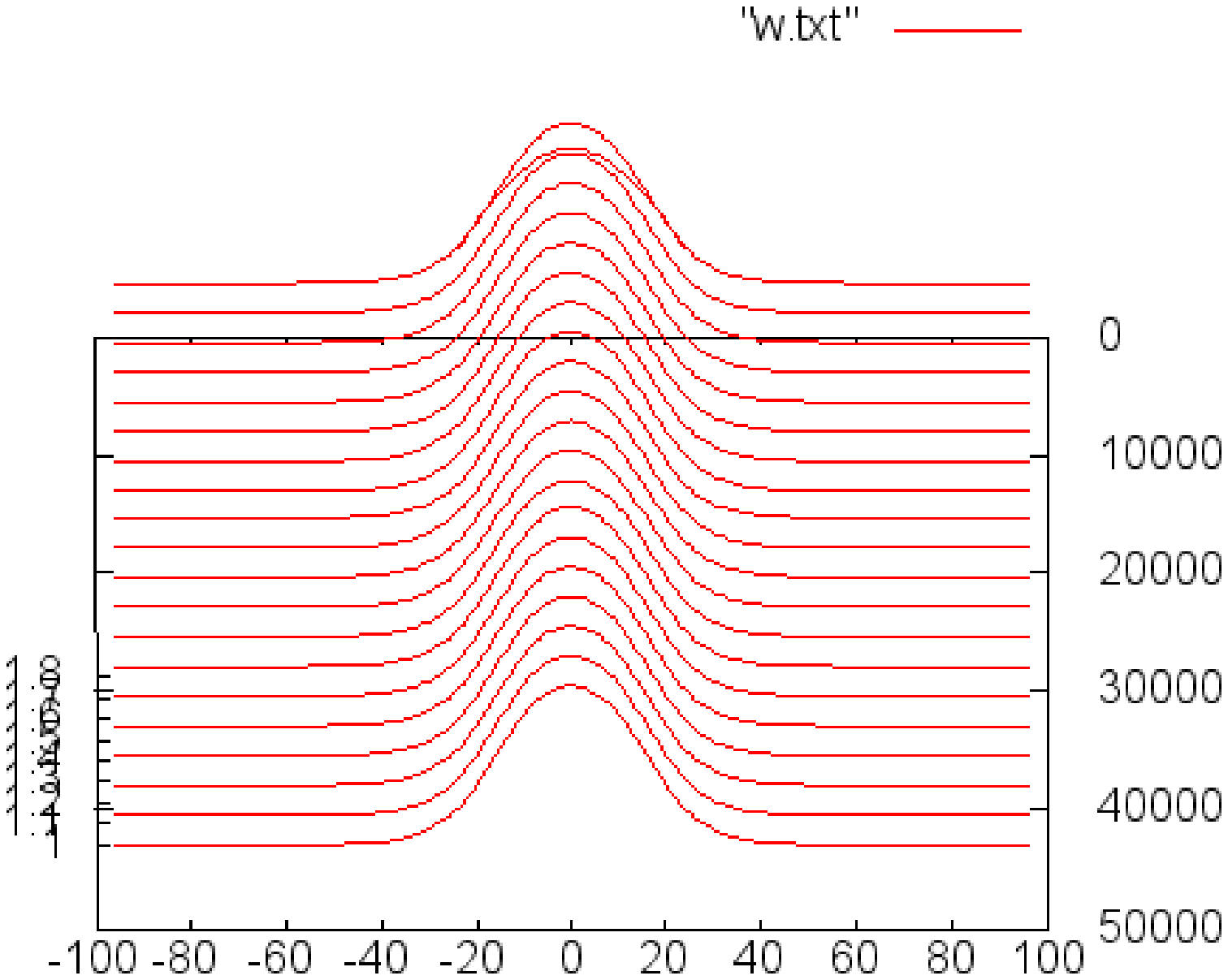}
\end{center}
\caption{$a=0.08,\ b=0.17$}
\label{6}
\end{minipage}
\end{figure}
In Figure \ref{5} and Figure \ref{6}, two types of steady state is observed.
\begin{figure}[H]
\begin{center}
\includegraphics[scale=0.3]{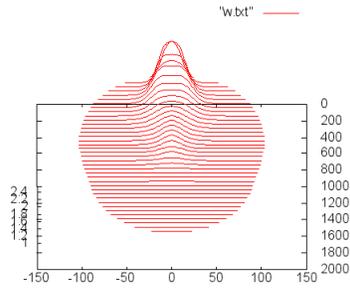}
\end{center}
\caption{$a=0.05,\ b=0.15$}
\label{7}
\end{figure}
In Figure \ref{7},\ values of $(u,w)$ converge to the stable equilibrium point $P_{d,0}$.

\begin{figure}[H]
\begin{center}
\includegraphics[scale=0.5]{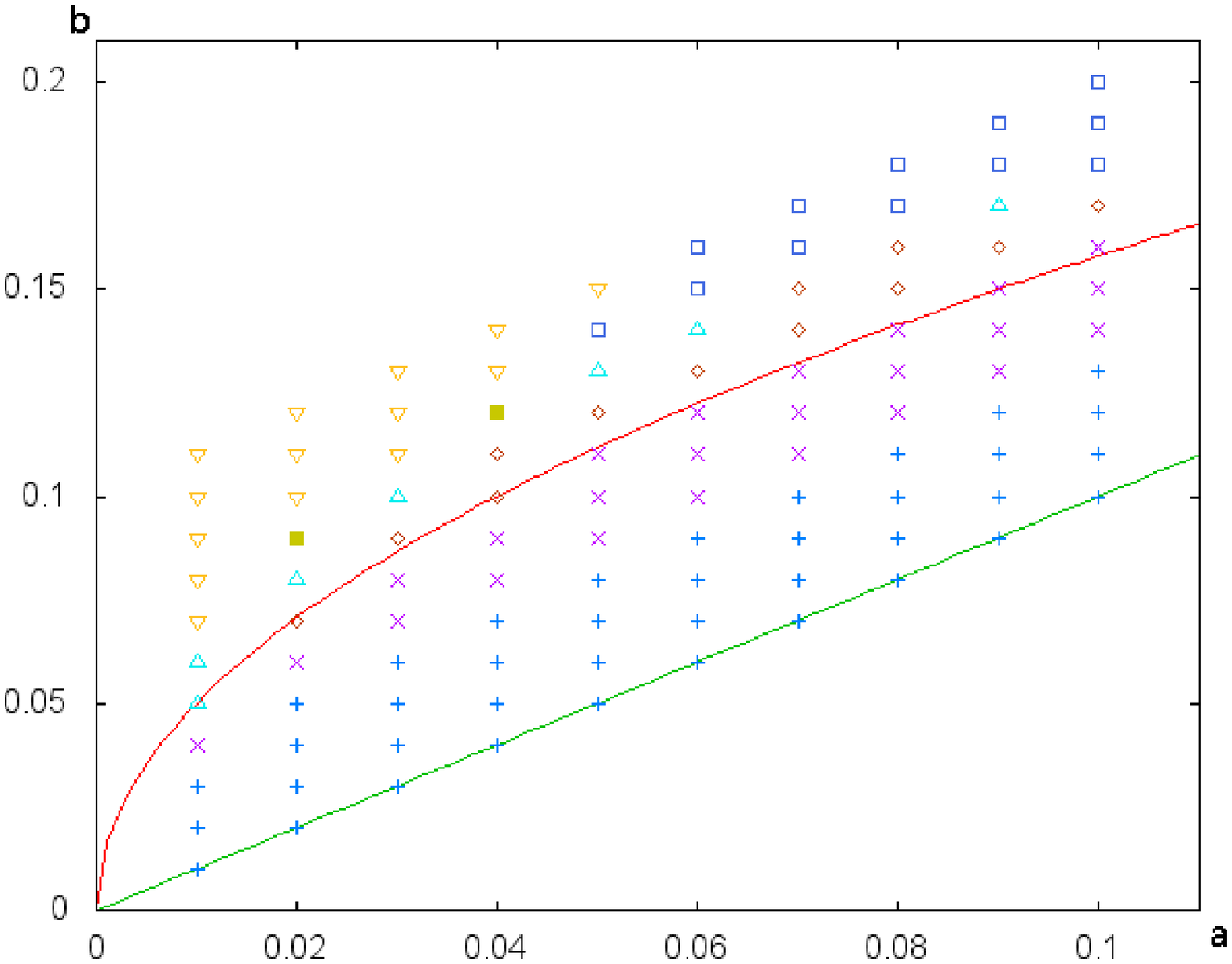}
\caption{}
\label{phase}
\end{center}
\end{figure}
Figure \ref{phase} reveals that which patterns are seen with the parameters $(a,b)$.
The horizontal axis is for $a$ and the vertical axis for $b$.
The upper curve is $a=4b^2$ and the lower line is $b=a$.
$\triangle$: Figure \ref{1}; +: Figure \ref{2}; $\times$: Figure \ref{3};$\diamond$: Figure \ref{4}; $\blacksquare$: Figure \ref{5}; $\square$: Figure \ref{6}; $\nabla$: Figure \ref{7}.
\section{Ultradiscrete Gray-Scott model}
\label{s3}
In this section, we ultradiscretize \eqref{dgs} and investigate the solutions.\\
Let
\begin{eqnarray*}
u_n^{\vec{j}}=\exp{\left(\frac{U_n^{\vec{j}}}{\varepsilon}\right)},\ w_n^{\vec{j}}=\exp{\left(\frac{W_n^{\vec{j}}}{\varepsilon}\right)},\\
\delta=\exp{\left(\frac{D}{\varepsilon}\right)},\ a=\exp{\left(\frac{A}{\varepsilon}\right)},\ b=\exp{\left(\frac{B}{\varepsilon}\right)}
\end{eqnarray*}
and take the limit $\varepsilon\to0$, then we have
\begin{equation}\label{udgs}
\begin{cases}
U_{n+1}^{\vec{j}}=\max{[M_p(U_n^{\vec{j}}),D+\max{[M_p(U_n^{\vec{j}})+W_{n+1}^{\vec{j}},A]}]}\\
~~~~~~~~~~~~~~~~~~~~~~~~~~~~~~~~-\max{[0,D+\max{[2W_{n+1}^{\vec{j}},A,0]}]},\\
W_{n+1}^{\vec{j}}=\max{[M_q(W_n^{\vec{j}}),D+\max{[M_p(U_n^{\vec{j}})+\max{[2M_q(W_n^{\vec{j}}),0]},B]}]}\\
~~~~~~~~~~~~~~~~~~~~~~~~~~~~~~~~-\max{[0,D+\max{[M_p(U_n^{\vec{j}}),B]}]},
\end{cases}
\end{equation}
where
\begin{equation*}
M_p(F^{\vec{j}}):=\max\limits_{k=1,\dots,d}{[F^{\vec{j}+p\vec{e}_k},F^{\vec{j}-p\vec{e}_k}]}.
\end{equation*}
Taking a limit $D\to\infty$ and assuming $W_n^{\vec{j}}\ge0$, then we get
\begin{equation}\label{udgs2}
\begin{cases}
U_{n+1}^{\vec{j}}=\max{[M_p(U_n^{\vec{j}})+W_{n+1}^{\vec{j}},A]-\max{[2W_{n+1}^{\vec{j}},A]}},\\
W_{n+1}^{\vec{j}}=\max{[M_p(U_n^{\vec{j}})+2M_q(W_n^{\vec{j}}),B]}-\max{[M_p(U_n^{\vec{j}}),B]}.
\end{cases}
\end{equation}
Let $d=1$ and initial data of \eqref{udgs2} $-U_0^j\in\{0,1\},W_0^j\in\{0,1\}$.
Taking some conditions to parameters $A$ and $B$, the solution of \eqref{udgs2} becomes to a cellular automaton.
There are several types of conditions for $A$ and $B$ as follow:
\begin{center}
\begin{tabular}{c|c|c|c|c}
Type I&Type II&Type III&Type IV&Type V\\
\hline
$A\le -1$&$0\le A\le 1$&$A\ge 2$&$A \le -1$&$A \ge 0$\\
$B=1$&$B=1$& $B=1$&$B\ge 2$&$B\ge 2$
\end{tabular}
\end{center}

Type I: The rule for $A\le -1,B=1$:
\begin{equation*}
\begin{array}{c|c|c|c|c}
-M_p(U_n^j),M_q(W_n^j)&1,1&1,0&0,1&0,0\\
\hline
-U_{n+1}^j,W_{n+1}^j&1,0&1,0&1,1&0,0
\end{array}
\end{equation*}
In this case, moving pulses are observed.
If two pulses collide, each pulse is disappeared.
\begin{figure}[htbp]
 \begin{center}
  \includegraphics[width=43mm]{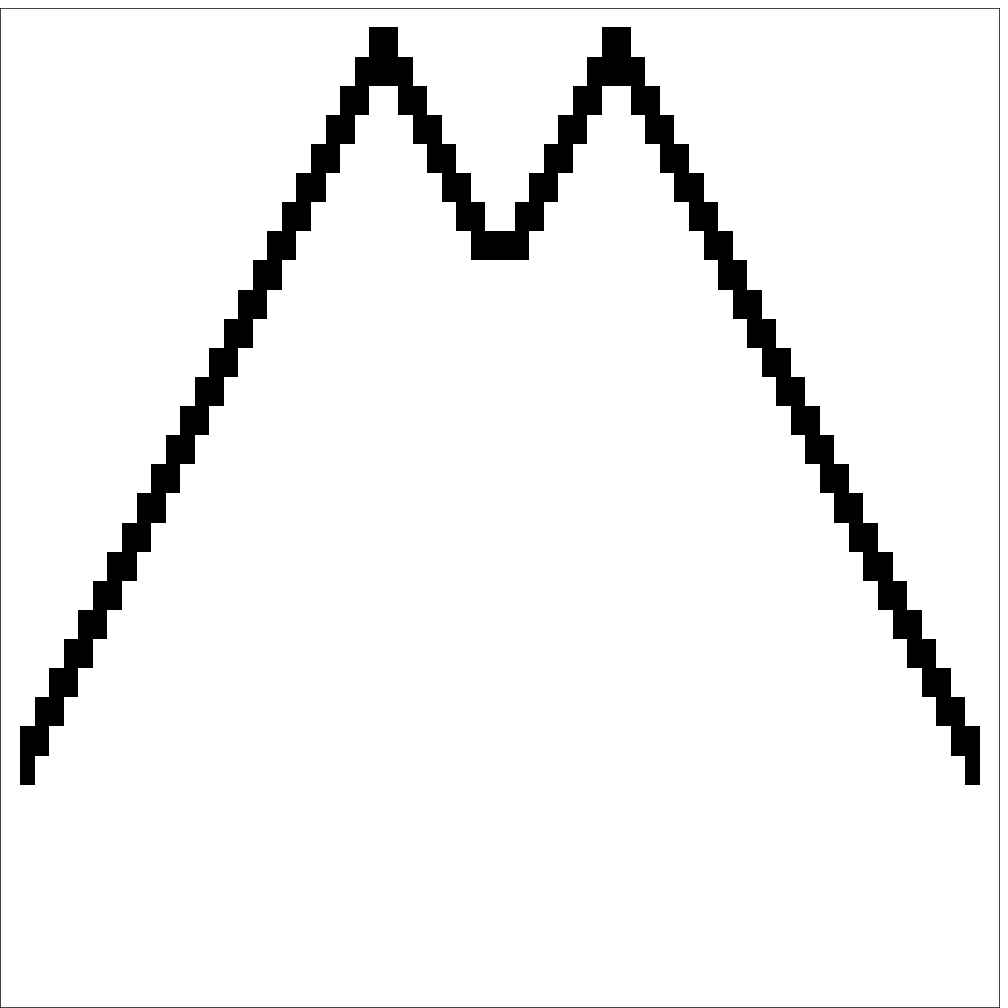} \qquad
   \includegraphics[width=43mm]{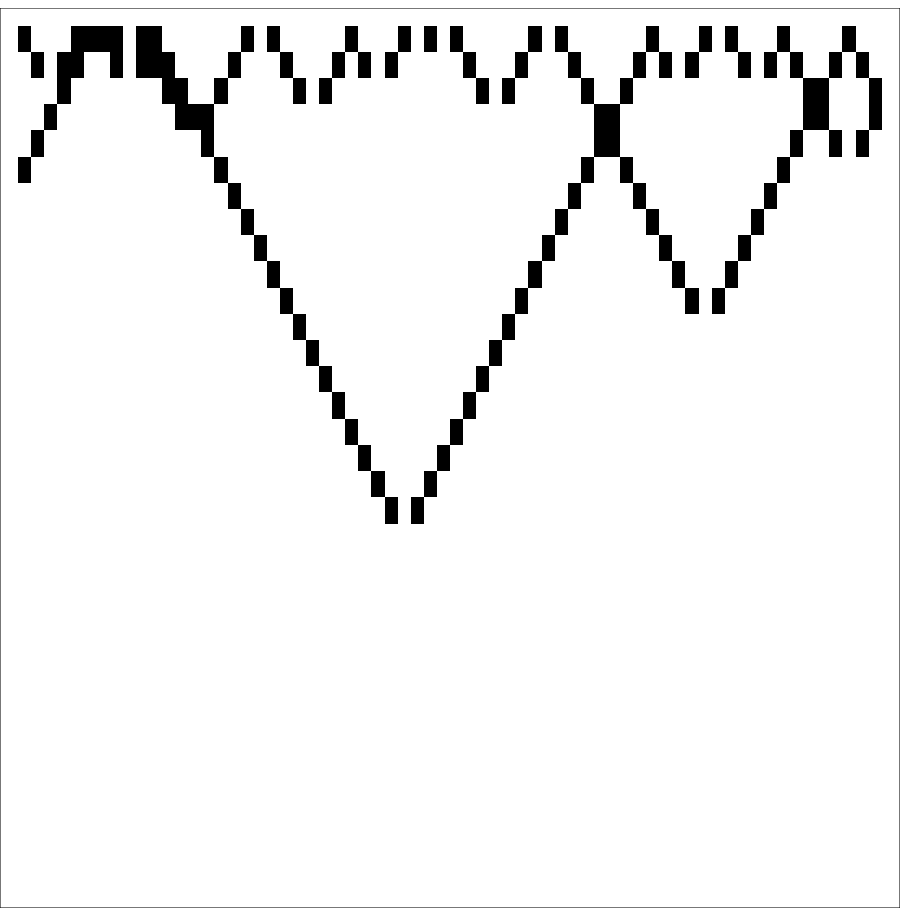}
\end{center}
 \caption{ $W_n^j$ with $ A = -1$, $B=1$ and $p=q=1$.}
 \label{fig:annihilate}
\end{figure}
\\
Values of $W_n^j$ is represent as follow: 0 (white) and 1 (black).

Type II: The rule for $0\le A \le 1,B=1$:
\begin{equation*}
\begin{array}{c|c|c|c|c}
-M_p(U_n^{j}),M_q(W_n^{j})&1,1&1,0&0,1&0,0\\
\hline
-U_{n+1}^j,W_{n+1}^j&0,0&0,0&1,1&0,0
\end{array}
\end{equation*}
In this case,\ $U_{n+1}^j=-W_{n+1}^j$.
Since this relation is held,\ $W_n^j$ satisfies a single equation.
Moreover, taking $p=q=1$, the equation is same as ECA rule 90, which is well known for fractal design:
\begin{equation*}
\begin{array}{c|c|c|c|c|c|c|c|c}
W_n^{j-1}\,W_n^j\,W_n^{j+1}&111&110&101&100&011&010&001&000\\
\hline
W_{n+1}^j&0&1&0&1&1&0&1&0
\end{array}
\end{equation*}
\begin{figure}[htbp]
 \begin{center}
  \includegraphics[width=43mm]{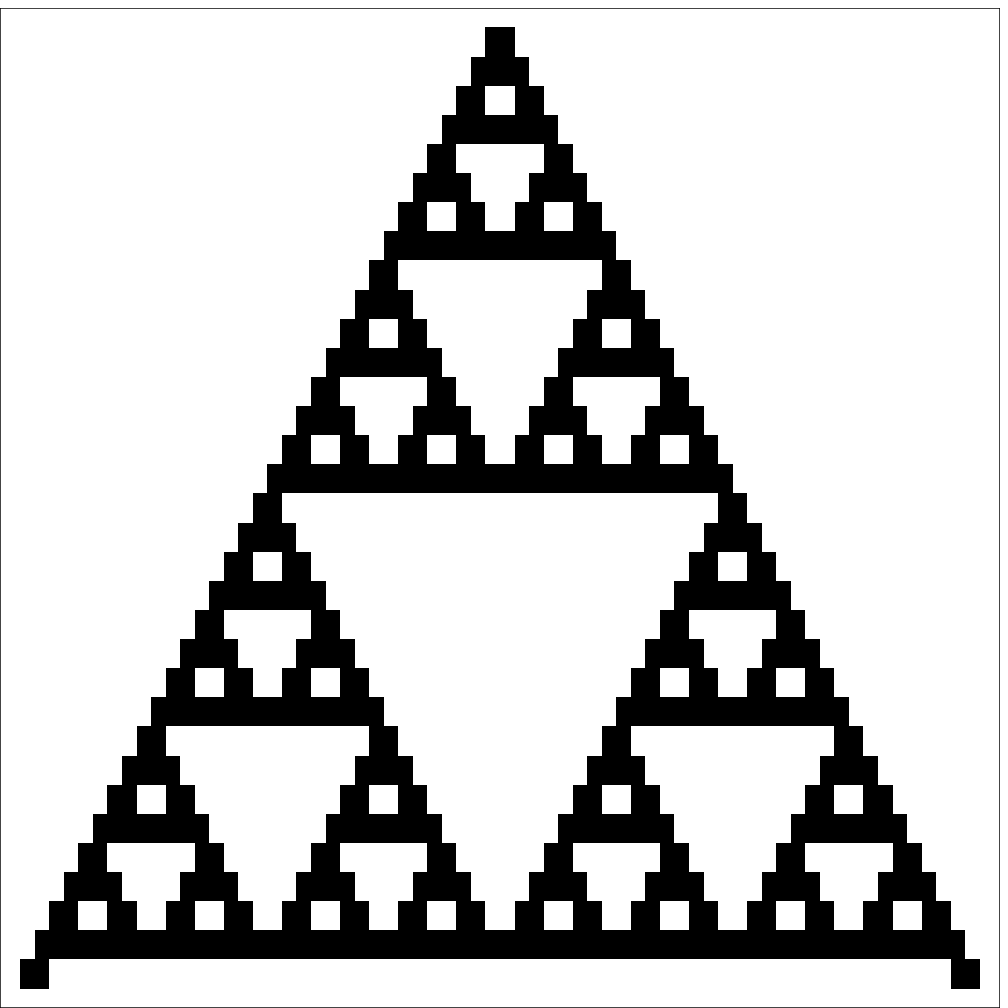}\qquad
  \includegraphics[width=43mm]{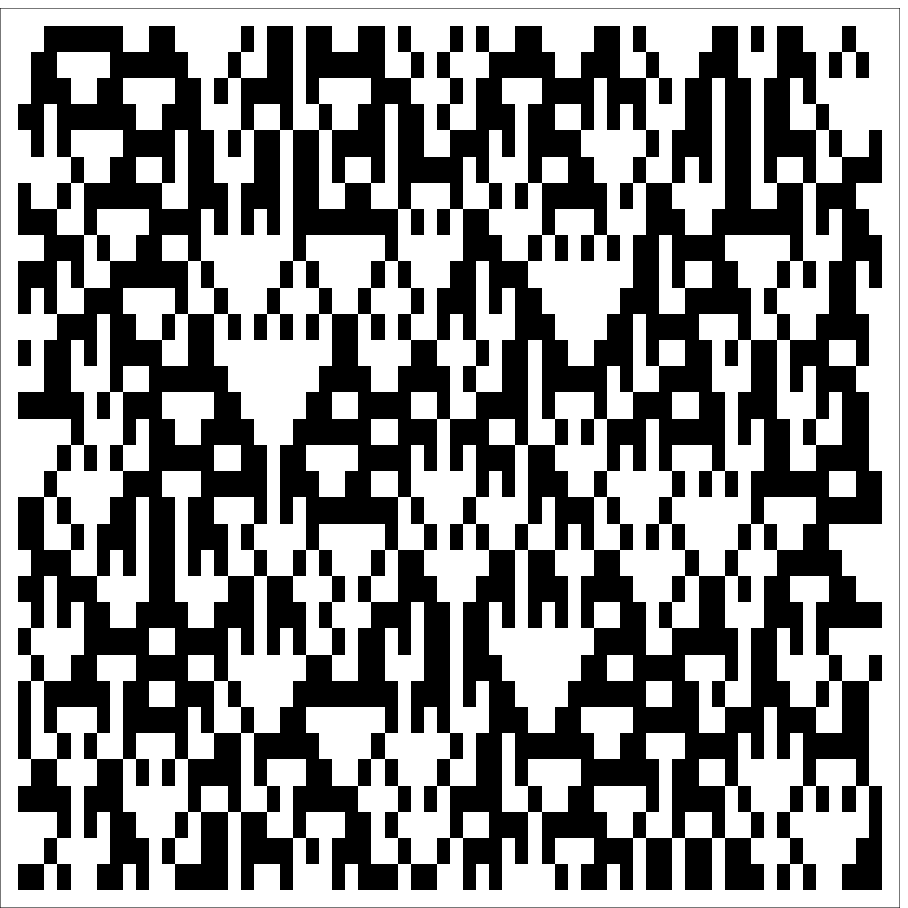}
 \end{center}
 \caption{$W_n^j$ with $ A =0,\,B=1$ and $p=q=1$.}
 \label{fig:eca90}
\end{figure}
\begin{figure}[htbp]
 \begin{center}
  \includegraphics[width=43mm]{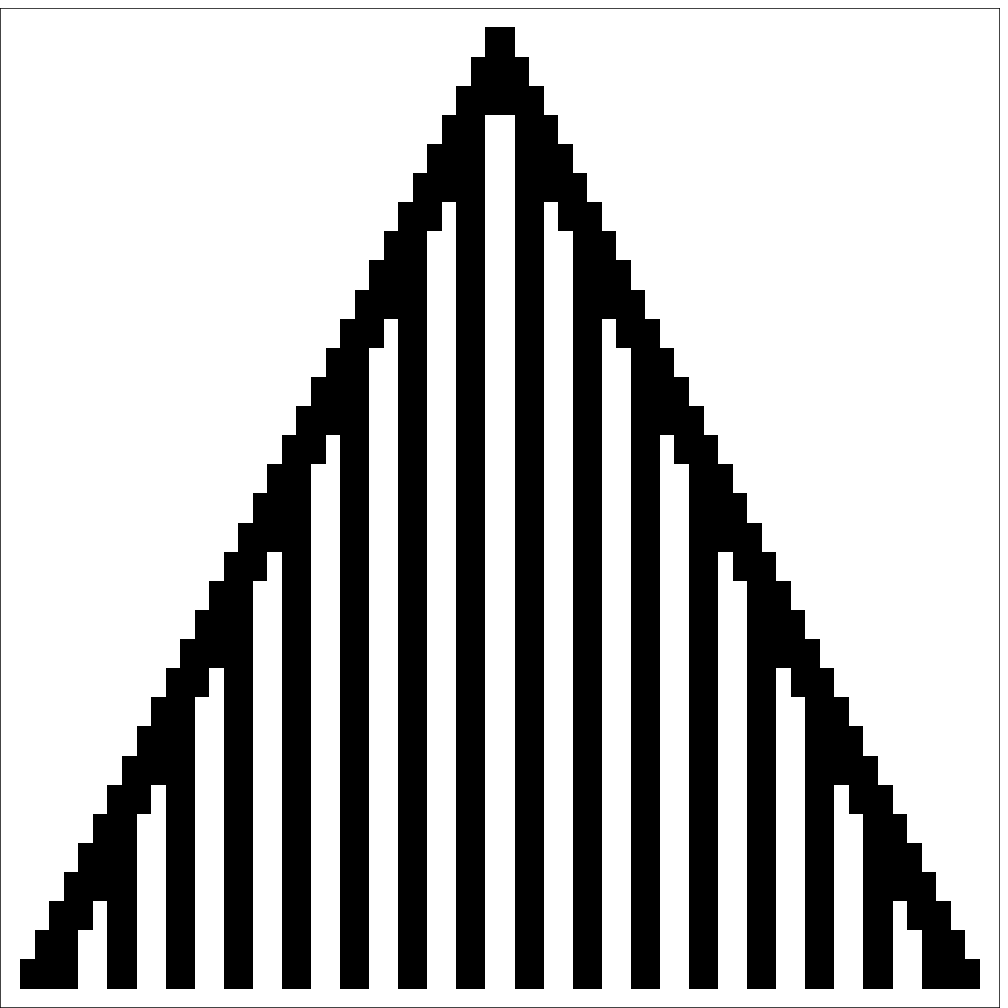}\qquad 
  \includegraphics[width=43mm]{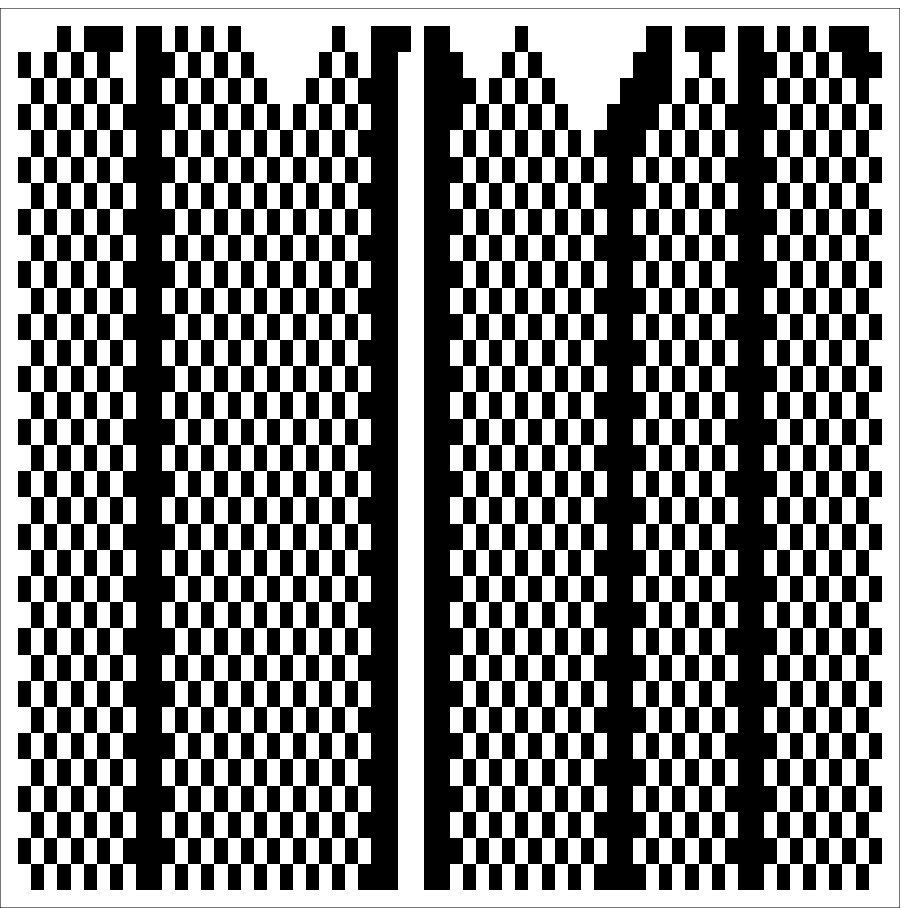}
 \end{center}
 \caption{$W_n^j$ with $A=0$,\ $B=1$,\ $p=2$ and $q=1$}
 \label{fig:self}
\end{figure}

Type III: The rule for $A\ge 2,B=1$:
\begin{equation*}
\begin{array}{c|c|c|c|c}
-M_p(U_n^j),M_q(W_n^j)&1,1&1,0&0,1&0,0\\
\hline
-U_{n+1}^j,W_{n+1}^j&0,0&0,0&0,1&0,0
\end{array}
\end{equation*}
In this case, $U_{n+1}^j=0$ so that $W_n^j$ satisfies $W_{n+1}^j=M_q(W_n^j)$.

Type IV: The rule of $A\le -1,B\ge 2$:
\begin{equation*}
\begin{array}{c|c|c|c|c}
-M_p(U_n^j),M_q(W_n^j)&1,1&1,0&0,1&0,0\\
\hline
-U_{n+1}^j,W_{n+1}^j&1,0&1,0&0,0&0,0
\end{array}
\end{equation*}
In this case, $W_{n+1}^j=0$ so that $U_n^j$ satisfies $U_{n+1}^j=M_p(U_n^j)$.

Type V: The rule of $A\ge 0,B\ge 2$:
\begin{equation*}
\begin{array}{c|c|c|c|c}
-M_p(U_n^j),M_q(W_n^j)&1,1&1,0&0,1&0,0\\
\hline
-U_{n+1}^j,W_{n+1}^j&0,0&0,0&0,0&0,0
\end{array}
\end{equation*}
In this case, $U_{n+1}^j=W_{n+1}^j=0$ so that $U$ and $W$ vanish immediately.

If one take $B\ge L,\ -U_n^j\in\{0,1,\dots,L\}$ and $W_n^j\in\{0,1,\dots,L\}$, the solution of \eqref{udgs2} becomes to a cellular automaton whose dependent variable can have $L+1$ values.
The more $L$ is large, the more the number of rule for evolution increases.
In this case, the spatial pattern is also classified five types as follow:
\begin{center}
\begin{tabular}{c|c|c|c|c}
Type I&Type II&Type III&Type IV&Type V\\
\hline
$A\le -1$&$0\le A\le 2L-1$&$A\ge 2L$&$A \le -1$&$A \ge 0$\\
$B=L$&$B=L$& $B=L$&$B\ge L+1$&$B\ge L+1$
\end{tabular}
\end{center}
In the case of type II, taking $L=2$ and $p=q=1$, a following Sierpinski gasket with shadow can be seen.
\begin{figure}[htbp]
 \begin{center}
  \includegraphics[width=43mm]{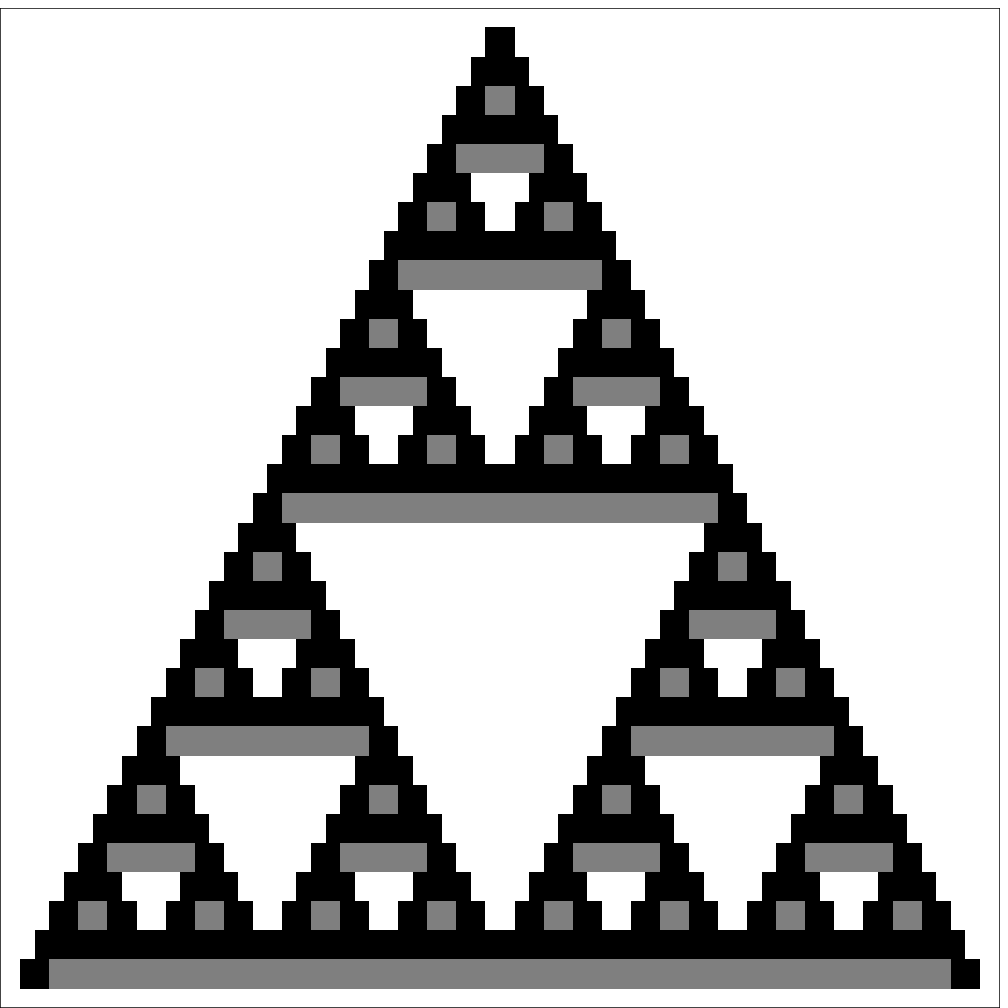}\qquad 
  \includegraphics[width=43mm]{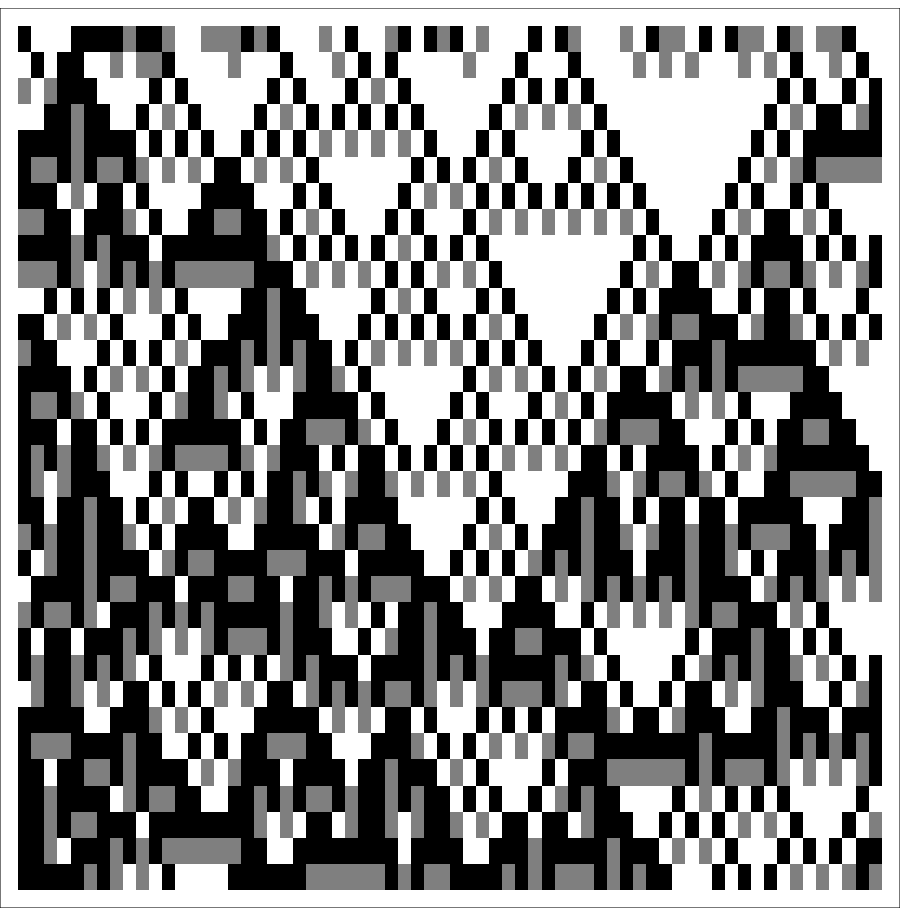}
 \end{center}
 \caption{$W_n^j\in\{0,1,2\}$ with $A=3$, $B=2$ and $p=q=1$}
 \label{fig:shade}
\end{figure}
\\
Values of $W_n^j$ is represent as follow: 0 (white), 1 (gray) and 2 (black).

Moreover, taking $p=2,\ q=1$, we can see the following patterns.
\begin{figure}[htbp]
 \begin{center}
  \includegraphics[width=43mm]{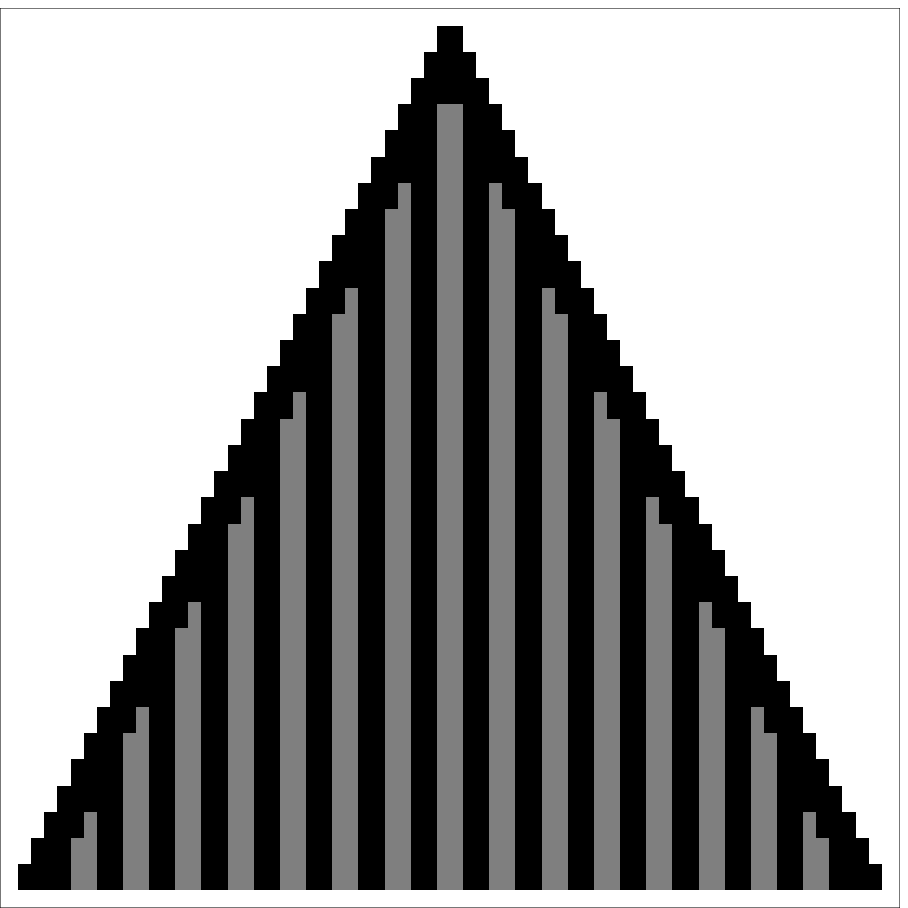}\qquad 
  \includegraphics[width=43mm]{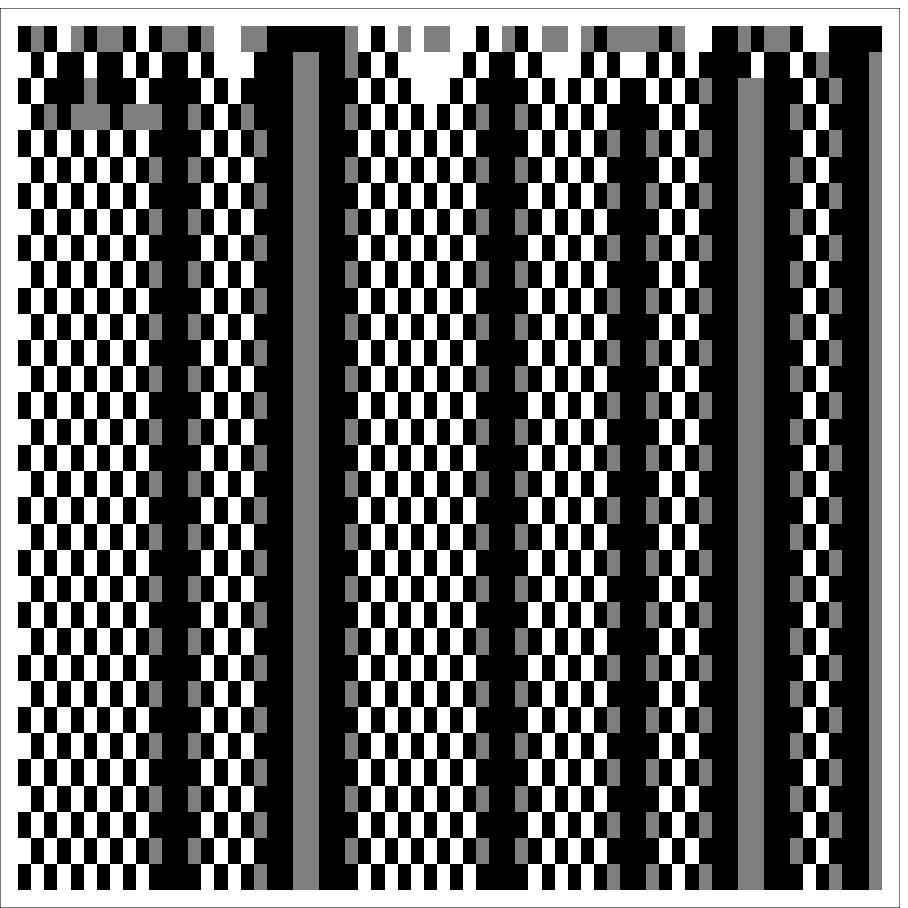}
 \end{center}
 \caption{$W_n^j\in\{0,1,2\}$ with $A=3$, $B=2$, $p=2$ and $q=1$}
 \label{fig:self2}
\end{figure}

Now, let $d=2$.
We also take similar condition to the initial condition of \eqref{udgs2} in the case of $d=1$:$W_0^{\vec{j}}\in\{0,1\}, -U_0^{\vec{j}}\in\{0,1\}.$
We can separate spatial patterns to five types as similar to the case of $d=1$.
If $A\le-1,B=1$ then the rule of the evolution is as follow:
\begin{equation*}
\begin{array}{c|c|c|c|c}
-M_p(U_n^{\vec{j}}),M_q(W_n^{\vec{j}})&1,1&1,0&0,1&0,0\\
\hline
-U_{n+1}^{\vec{j}},W_{n+1}^{\vec{j}}&1,0&1,0&1,1&0,0
\end{array}
\end{equation*}
In this case, the following pattern is observed.
\begin{figure}[htbp]
 \begin{minipage}{0.24\hsize}
  \begin{center}
   \includegraphics[width=29mm]{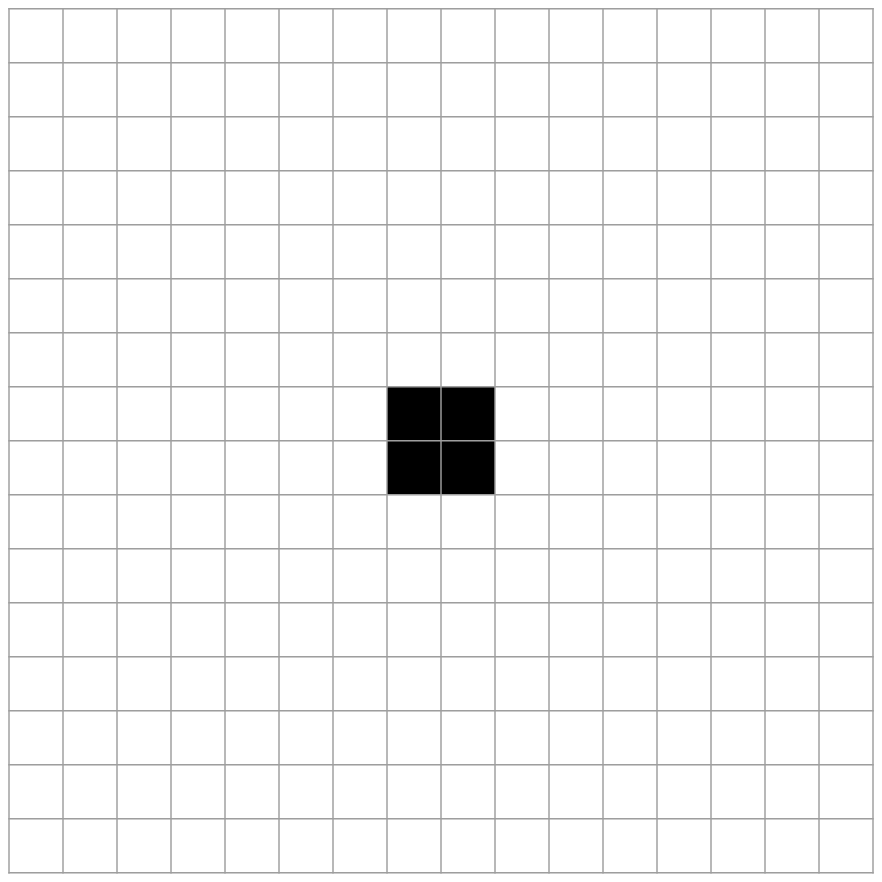}\\
   $n=0$
  \end{center}
  \label{fig:2dtravel-00}
 \end{minipage}
 \begin{minipage}{0.24\hsize}
  \begin{center}
   \includegraphics[width=29mm]{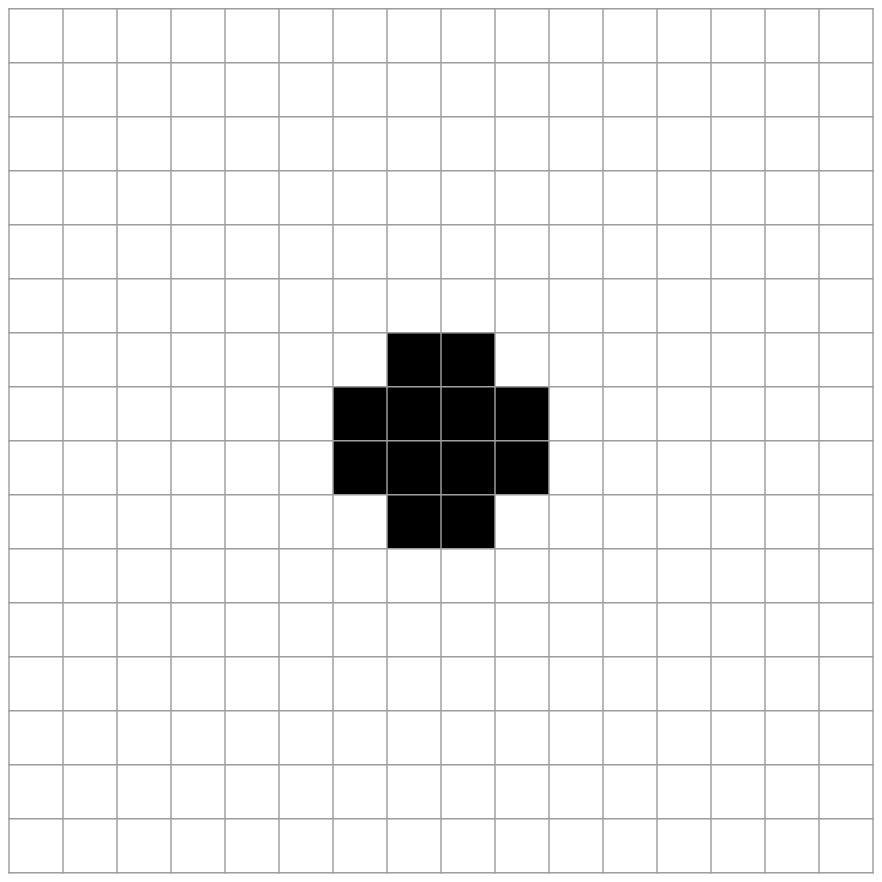}\\
   $n=1$
  \end{center}
  \label{fig:2dtravel-01}
 \end{minipage}
 \begin{minipage}{0.24\hsize}
  \begin{center}
   \includegraphics[width=29mm]{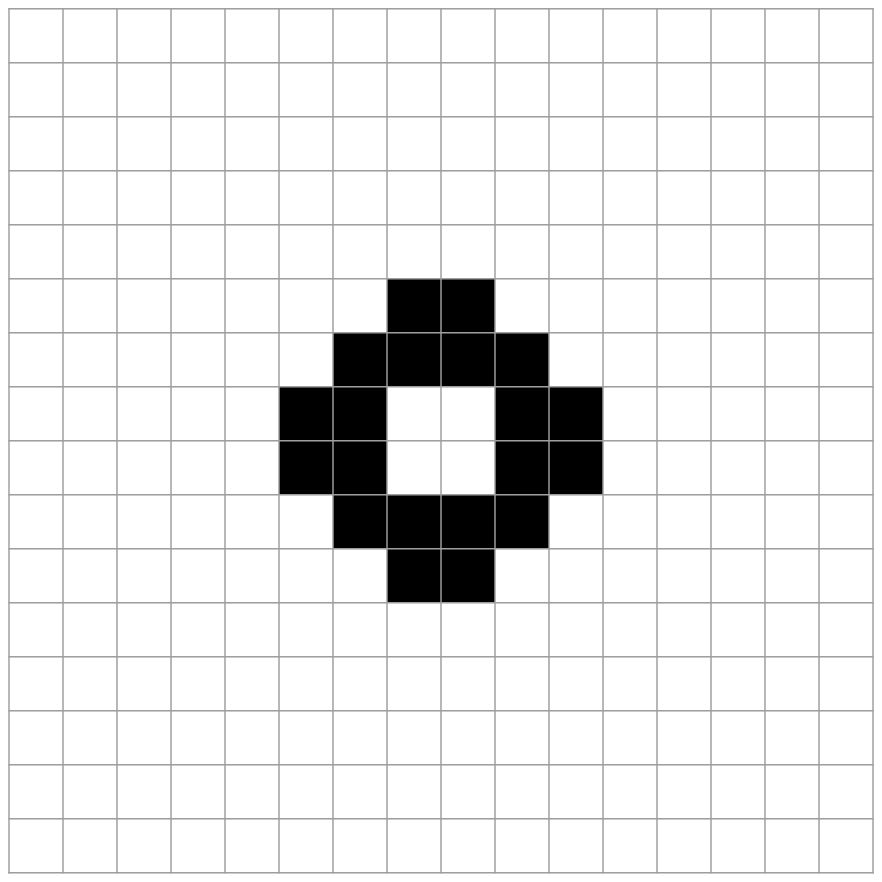}\\
   $n=2$
  \end{center}
  \label{fig:2dtravel-02}
 \end{minipage}
 \begin{minipage}{0.24\hsize}
  \begin{center}
   \includegraphics[width=29mm]{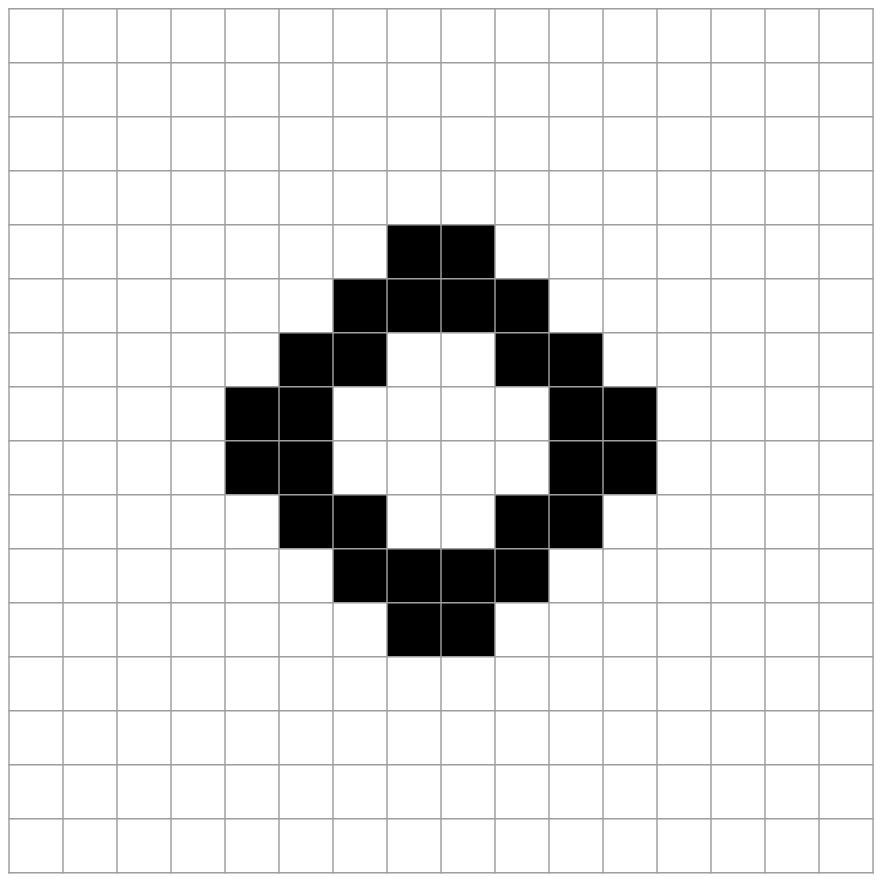}\\
   $n=3$
  \end{center}
  \label{fig:2dtravel-03}
 \end{minipage}\\
  \begin{minipage}{0.24\hsize}
  \begin{center}
   \includegraphics[width=29mm]{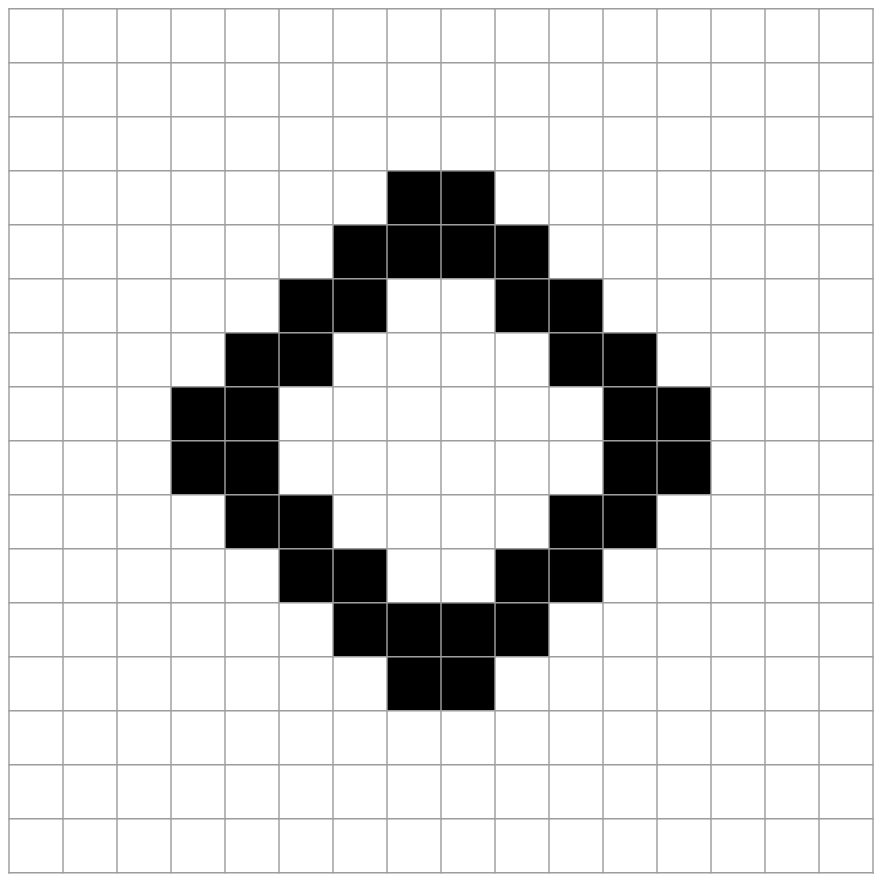}\\
   $n=4$
  \end{center}
  \label{fig:2dtravel-04}
 \end{minipage}
 \begin{minipage}{0.24\hsize}
  \begin{center}
   \includegraphics[width=29mm]{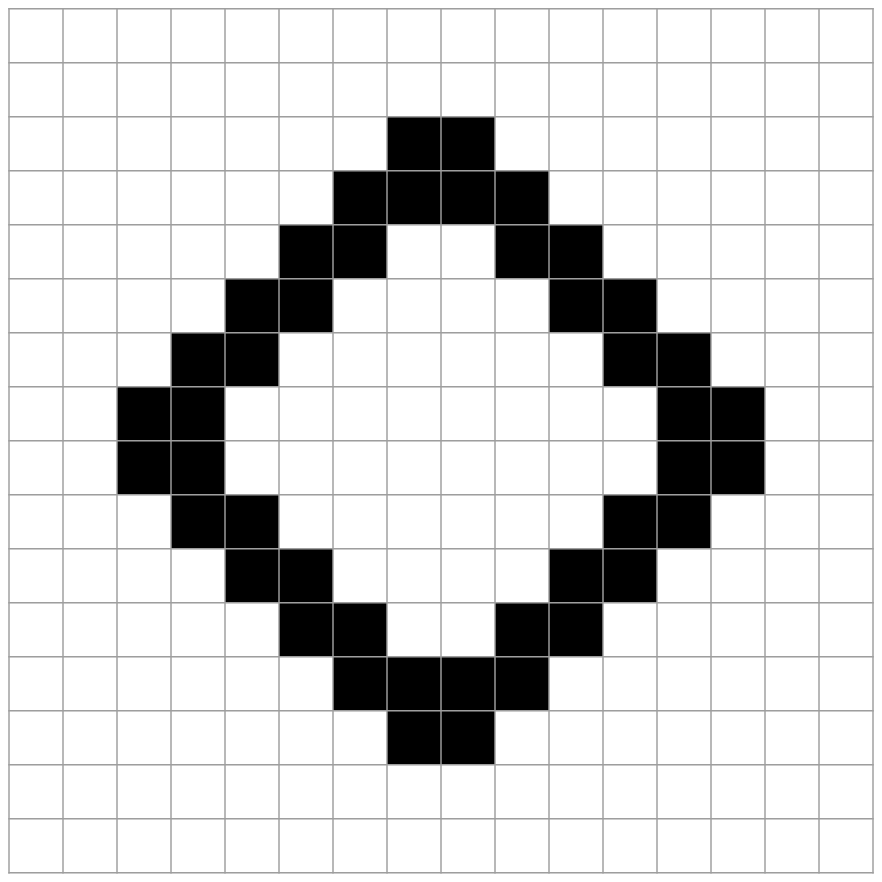}\\
   $n=5$
  \end{center}
  \label{fig:2dtravel-05}
 \end{minipage}
 \begin{minipage}{0.24\hsize}
  \begin{center}
   \includegraphics[width=29mm]{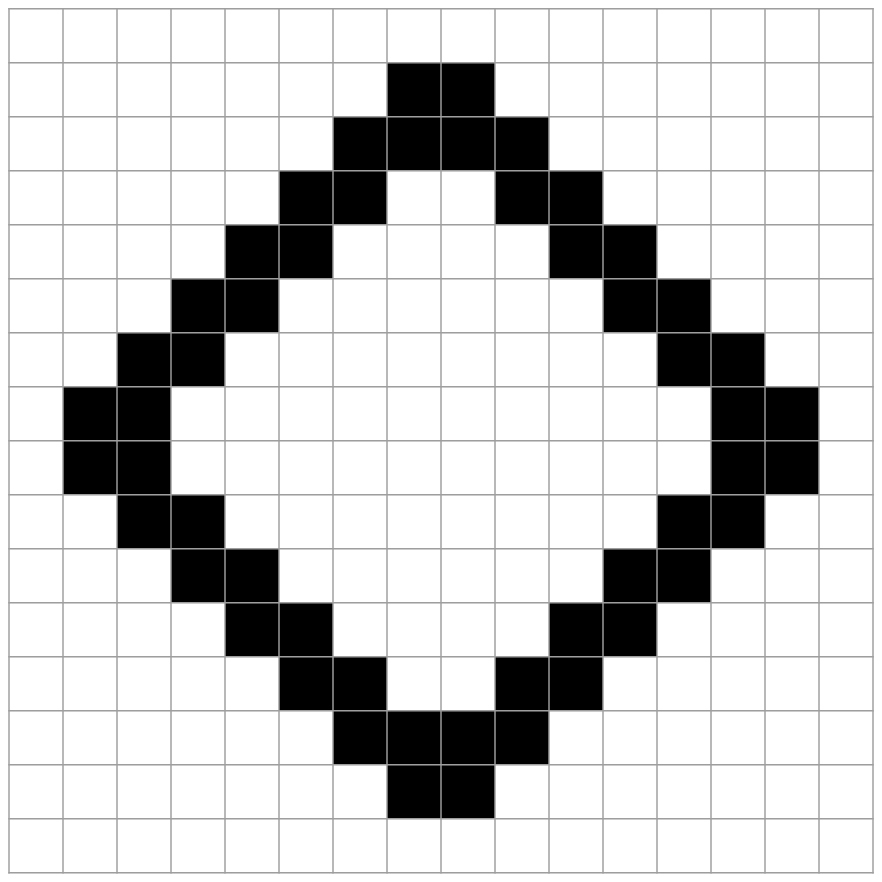}\\
   $n=6$
  \end{center}
  \label{fig:2dtravel-06}
 \end{minipage}
 \begin{minipage}{0.24\hsize}
  \begin{center}
   \includegraphics[width=29mm]{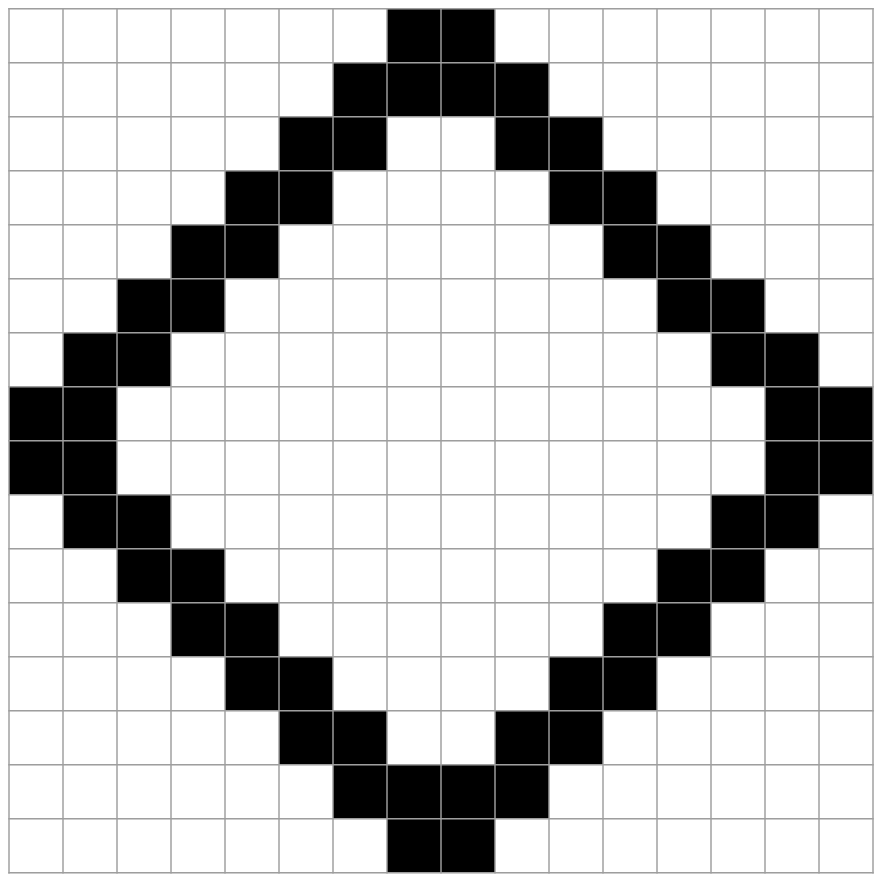}\\
   $n=7$
  \end{center}
  \label{fig:2dtravel-07}
 \end{minipage}\\
  \begin{minipage}{0.24\hsize}
  \begin{center}
   \includegraphics[width=29mm]{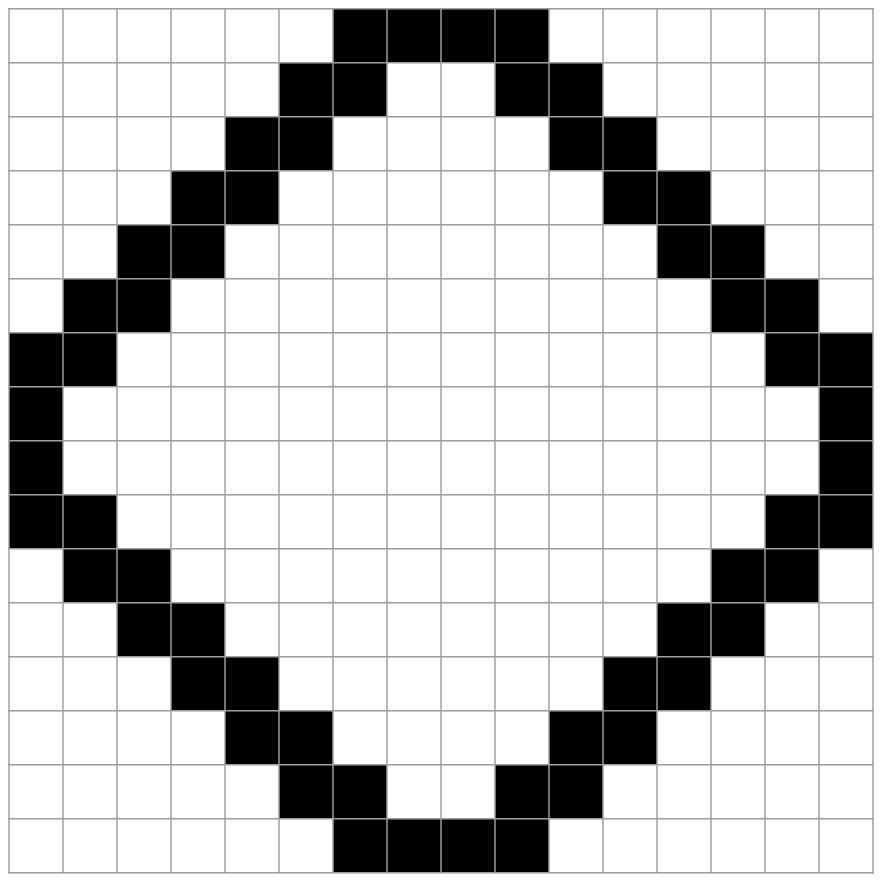}\\
   $n=8$
  \end{center}
  \label{fig:2dtravel-08}
 \end{minipage}
 \begin{minipage}{0.24\hsize}
  \begin{center}
   \includegraphics[width=29mm]{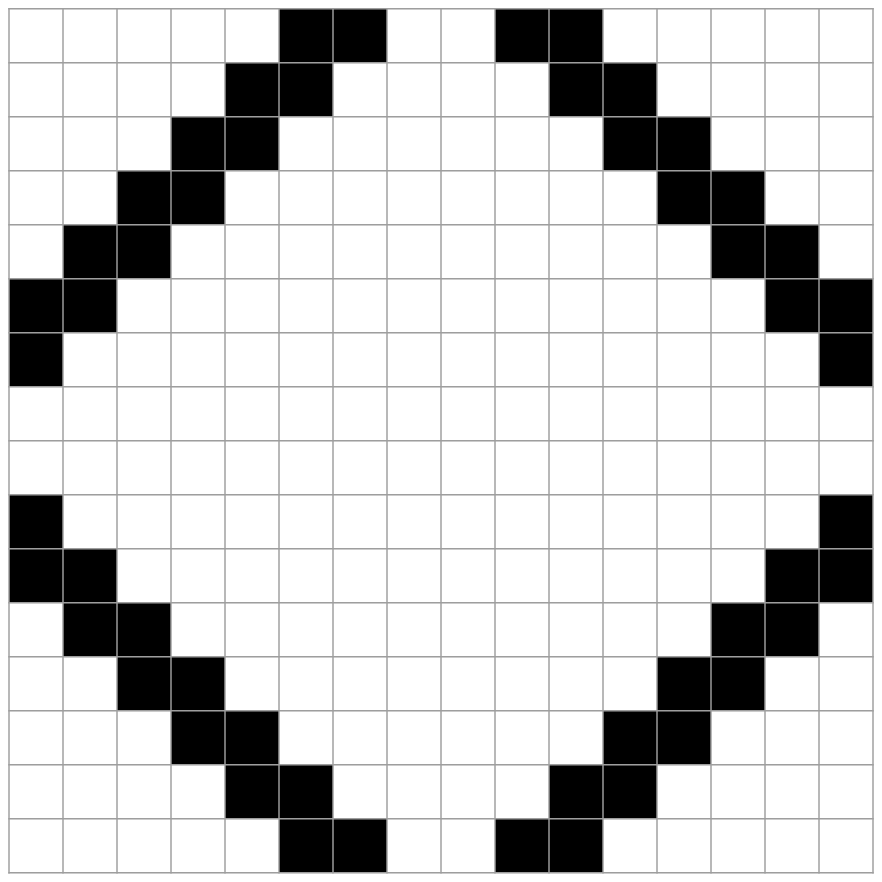}\\
   $n=9$
  \end{center}
  \label{fig:2dtravel-09}
 \end{minipage}
 \begin{minipage}{0.24\hsize}
  \begin{center}
   \includegraphics[width=29mm]{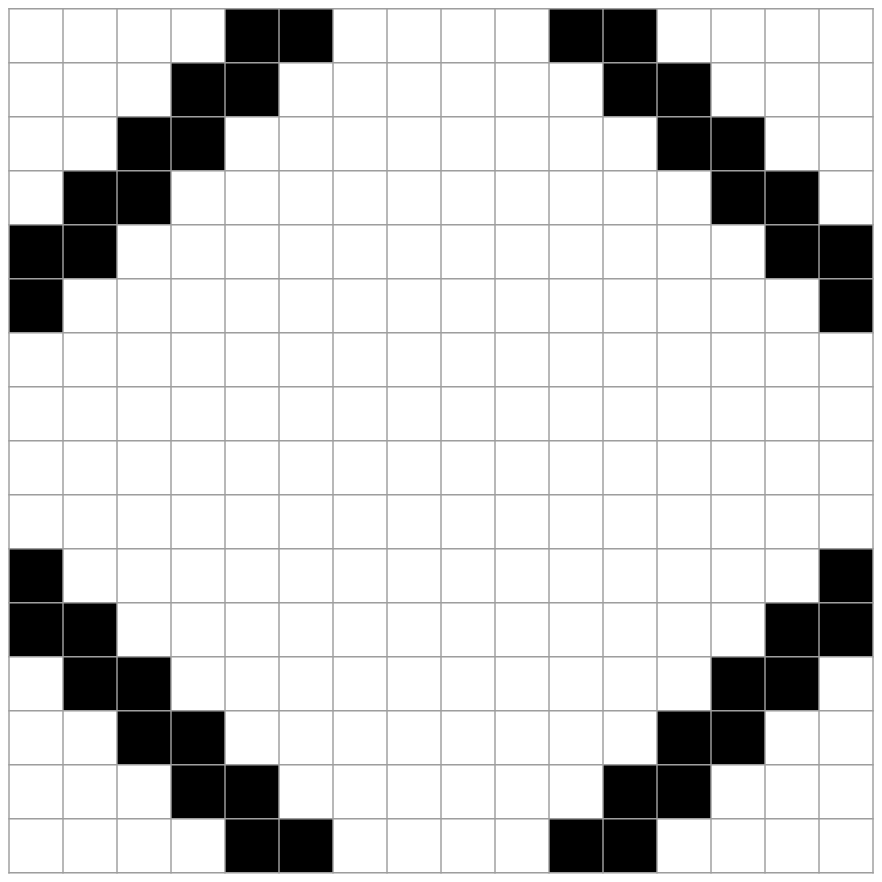}\\
   $n=10$
  \end{center}
  \label{fig:2dtravel-10}
 \end{minipage}
 \begin{minipage}{0.24\hsize}
  \begin{center}
   \includegraphics[width=29mm]{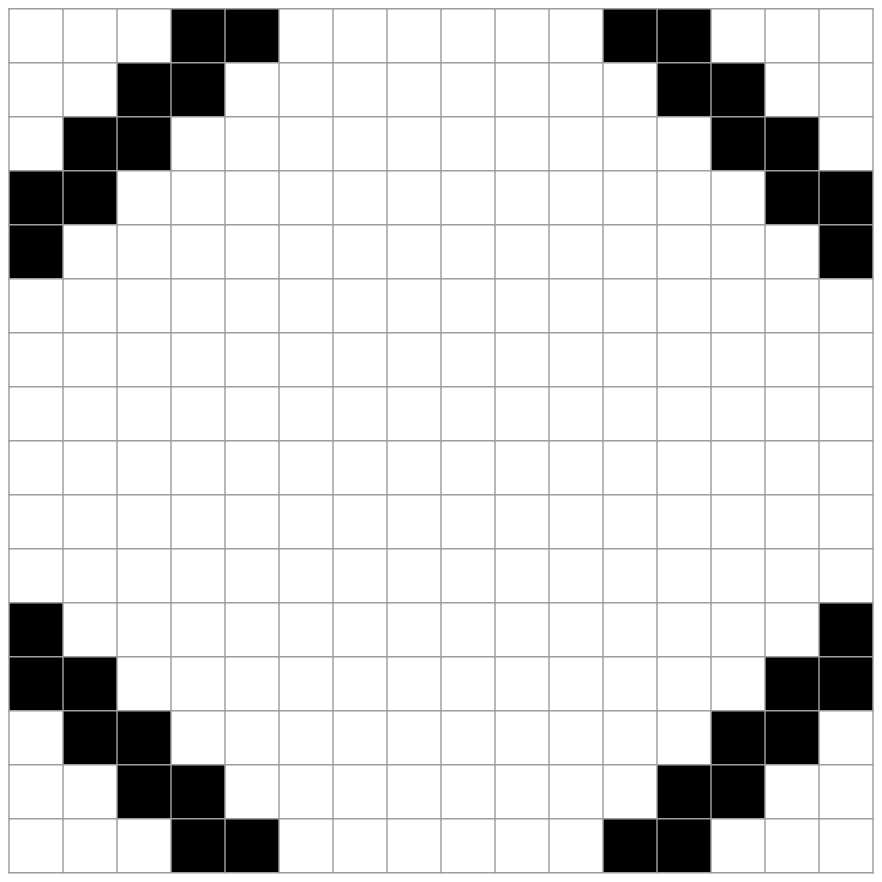}\\
   $n=11$
  \end{center}
  \label{fig:2dtravel-11}
 \end{minipage}
  \caption{A ring pattern is spreading.}\label{fig:2dtravel}
\end{figure}
Values of $W_n^{\vec{j}}$ is represent as follow: 0 (white) and 1 (black).

If $0\le A\le1,B=1$, the rule of evolution is as follow:
\begin{equation*}
\begin{array}{c|c|c|c|c}
-M_p(U_n^{\vec{j}}),M_q(W_n^{\vec{j}})&1,1&1,0&0,1&0,0\\
\hline
-U_{n+1}^{\vec{j}},W_{n+1}^{\vec{j}}&0,0&0,0&1,1&0,0
\end{array}
\end{equation*}
In this case, we can see the following patterns:
\begin{figure}[htbp]
 \begin{minipage}{0.24\hsize}
  \begin{center}
   \includegraphics[width=29mm]{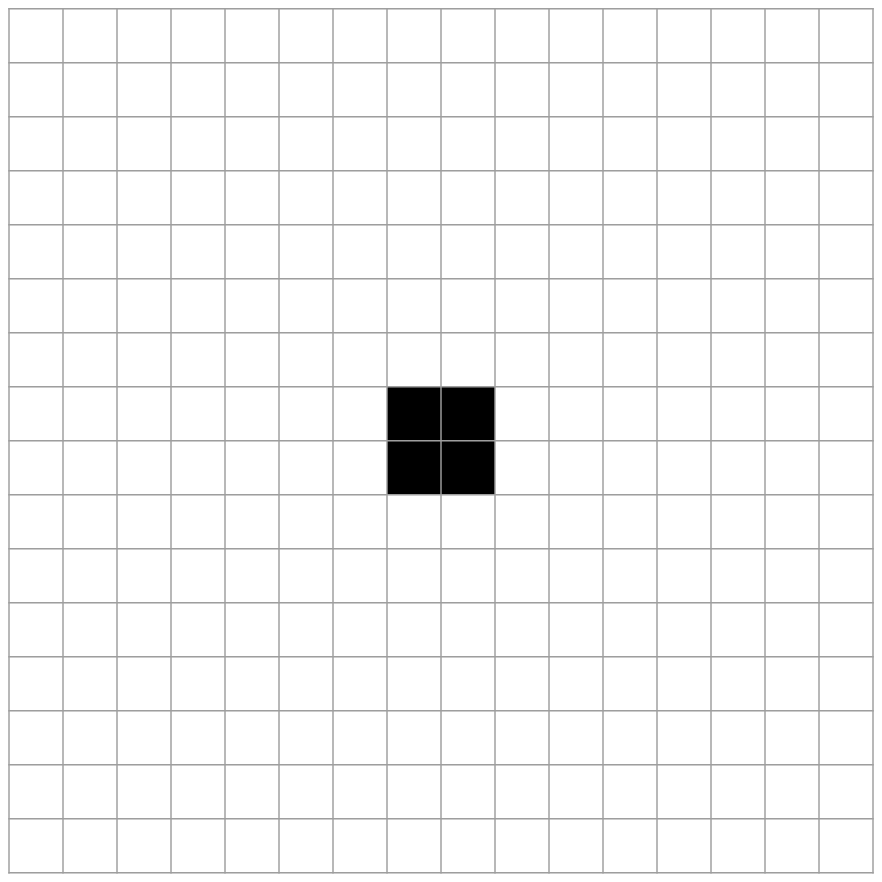}\\
   $n=0$
  \end{center}
  \label{fig:2dchaos-00}
 \end{minipage}
 \begin{minipage}{0.24\hsize}
  \begin{center}
   \includegraphics[width=29mm]{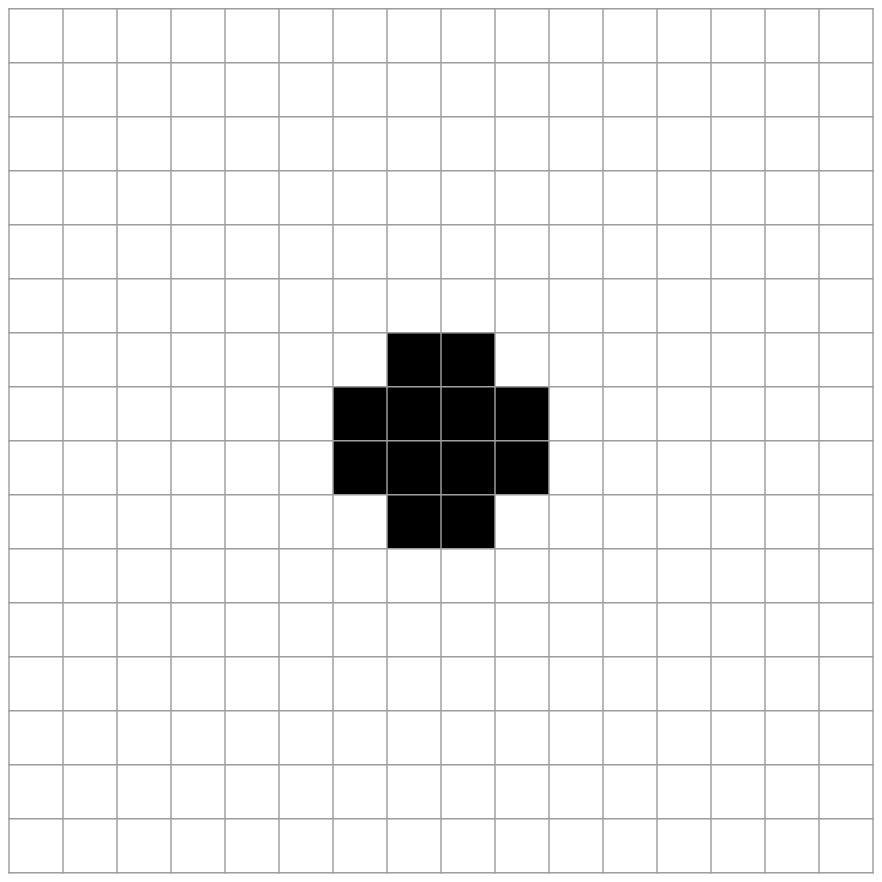}\\
   $n=1$
  \end{center}
  \label{fig:2dchaos-01}
 \end{minipage}
 \begin{minipage}{0.24\hsize}
  \begin{center}
   \includegraphics[width=29mm]{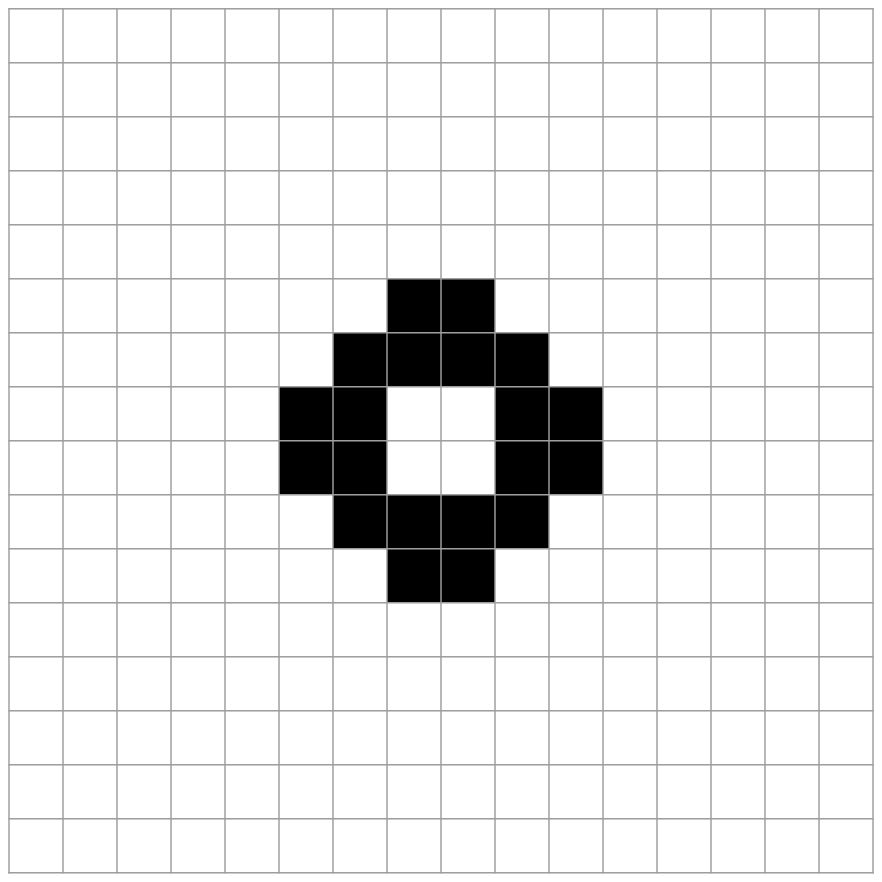}\\
   $n=2$
  \end{center}
  \label{fig:2dchaos-02}
 \end{minipage}
 \begin{minipage}{0.24\hsize}
  \begin{center}
   \includegraphics[width=29mm]{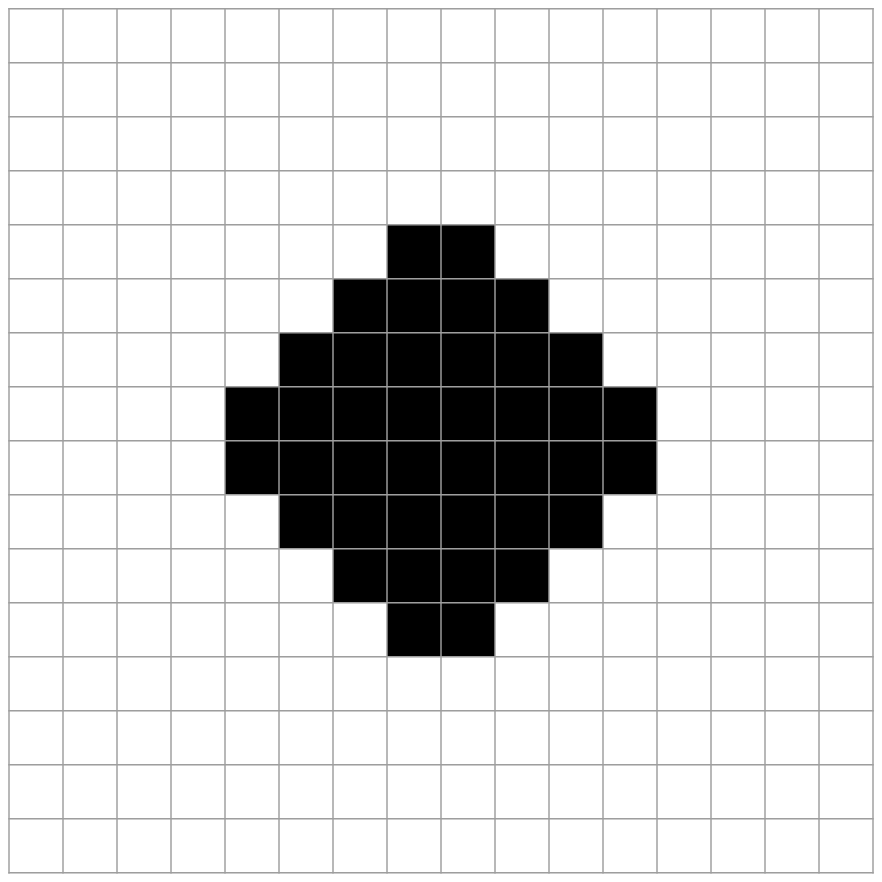}\\
   $n=3$
  \end{center}
  \label{fig:2dchaos-03}
 \end{minipage}\\
  \begin{minipage}{0.24\hsize}
  \begin{center}
   \includegraphics[width=29mm]{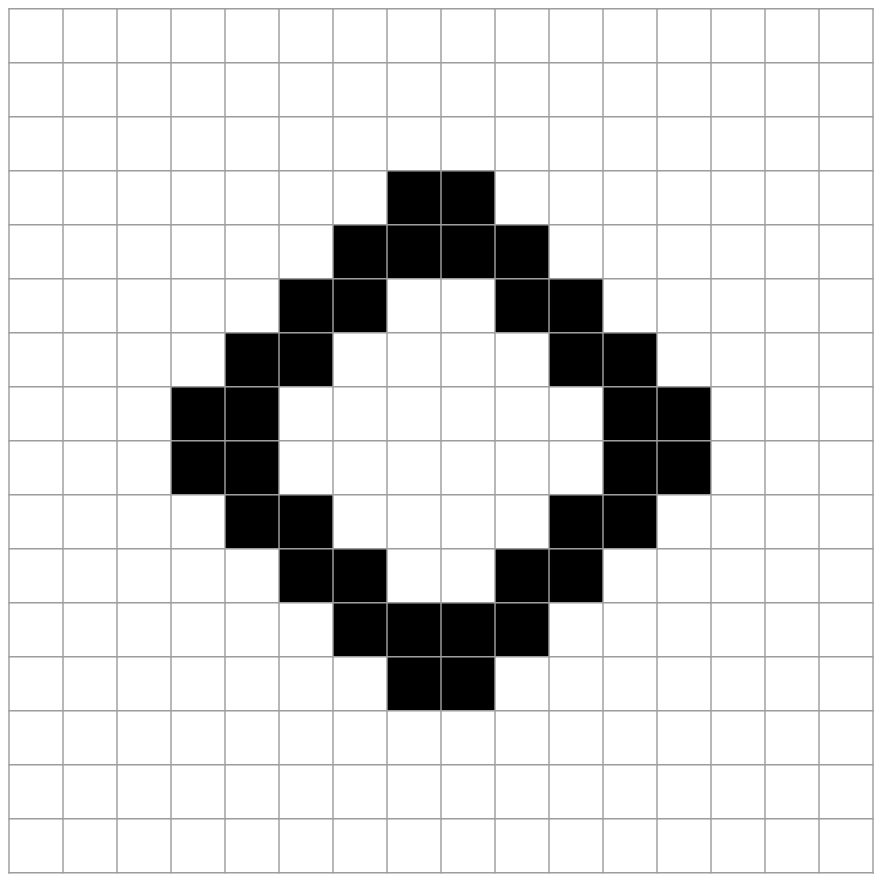}\\
   $n=4$
  \end{center}
  \label{fig:2dchaos-04}
 \end{minipage}
 \begin{minipage}{0.24\hsize}
  \begin{center}
   \includegraphics[width=29mm]{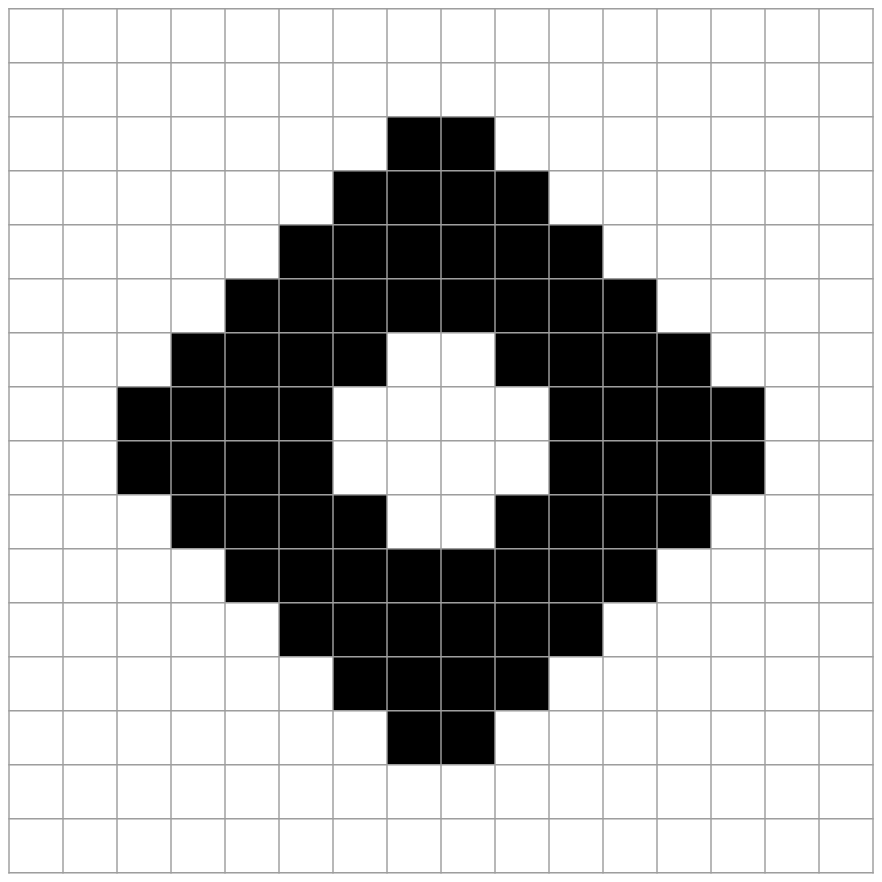}\\
   $n=5$
  \end{center}
  \label{fig:2dchaos-05}
 \end{minipage}
 \begin{minipage}{0.24\hsize}
  \begin{center}
   \includegraphics[width=29mm]{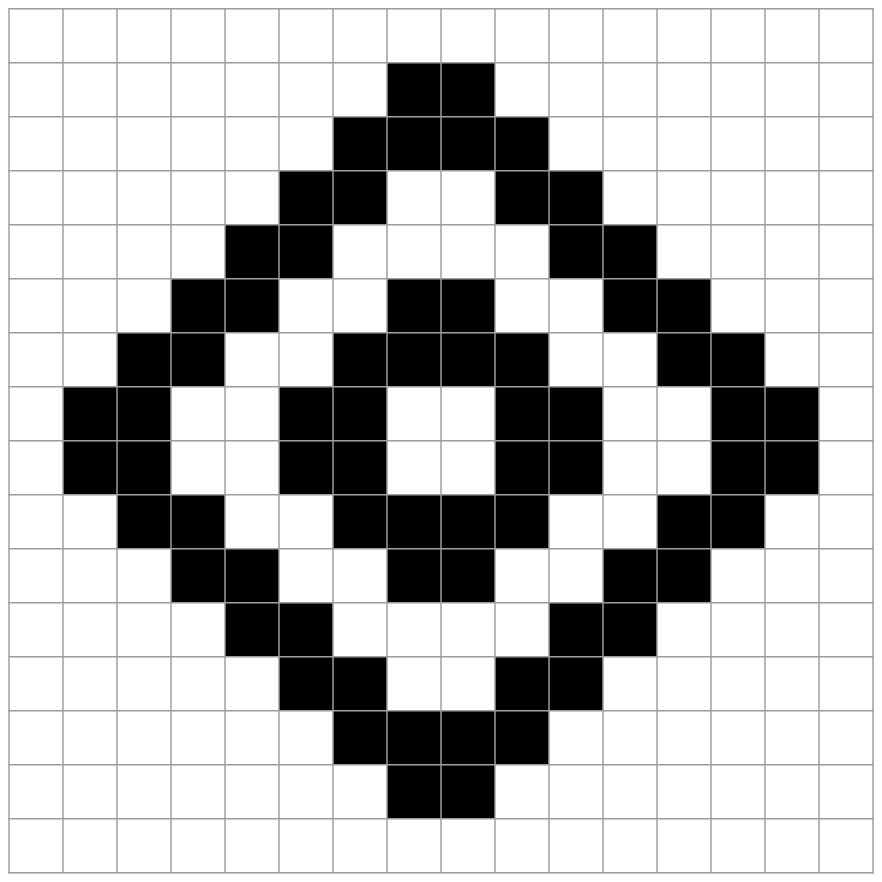}\\
   $n=6$
  \end{center}
  \label{fig:2dchaos-06}
 \end{minipage}
 \begin{minipage}{0.24\hsize}
  \begin{center}
   \includegraphics[width=29mm]{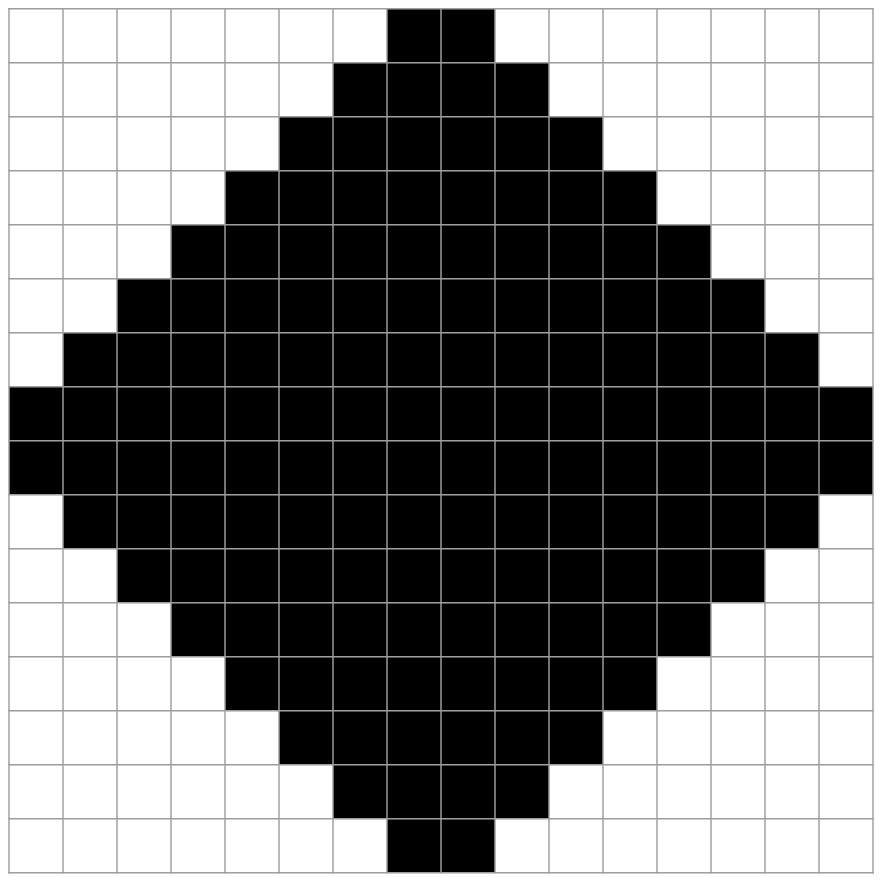}\\
   $n=7$
  \end{center}
  \label{fig:2dchaos-07}
 \end{minipage}\\
  \begin{minipage}{0.24\hsize}
  \begin{center}
   \includegraphics[width=29mm]{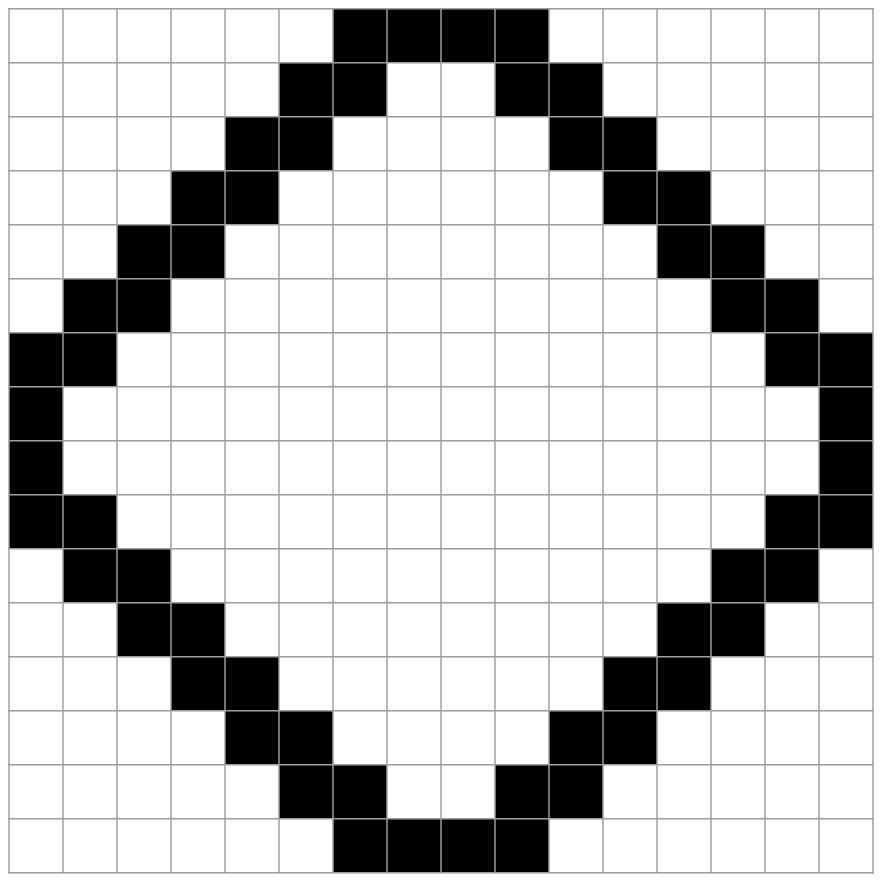}\\
   $n=8$
  \end{center}
  \label{fig:2dchaos-08}
 \end{minipage}
 \begin{minipage}{0.24\hsize}
  \begin{center}
   \includegraphics[width=29mm]{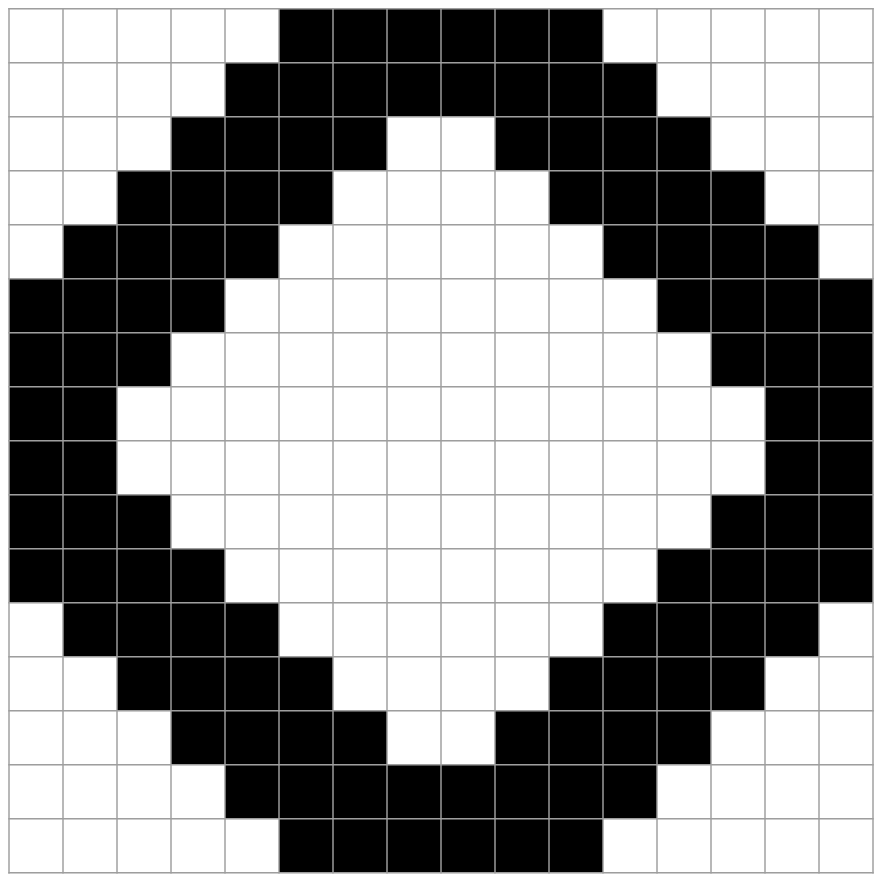}\\
   $n=9$
  \end{center}
  \label{fig:2dchaos-09}
 \end{minipage}
 \begin{minipage}{0.24\hsize}
  \begin{center}
   \includegraphics[width=29mm]{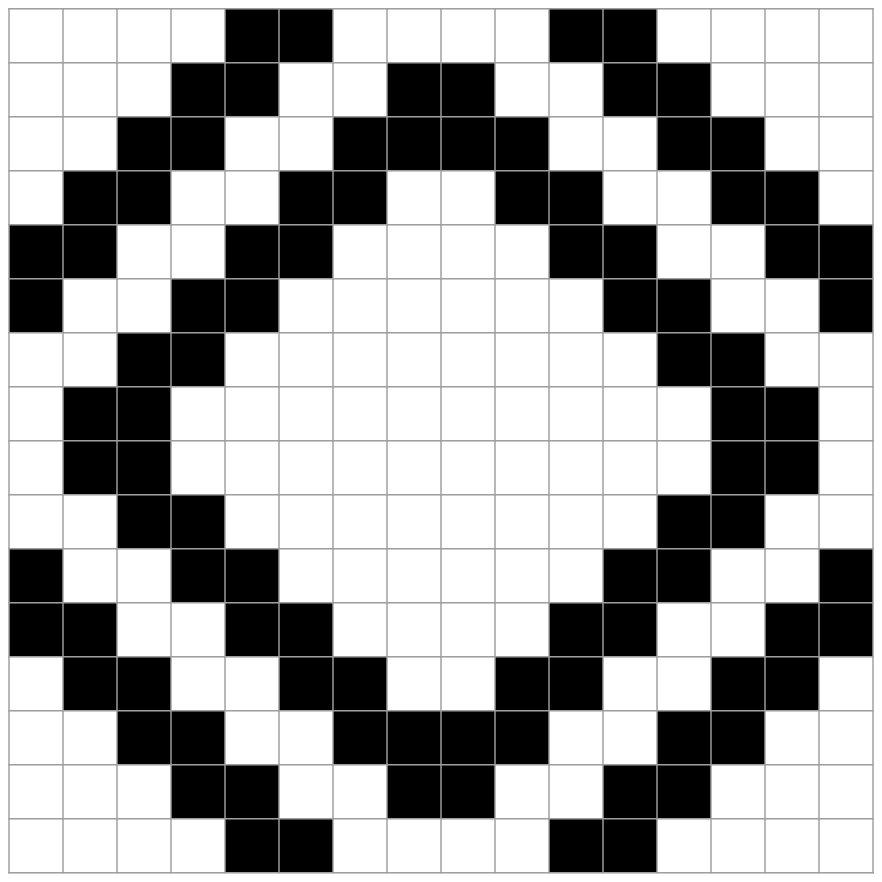}\\
   $n=10$
  \end{center}
  \label{fig:2dchaos-10}
 \end{minipage}
 \begin{minipage}{0.24\hsize}
  \begin{center}
   \includegraphics[width=29mm]{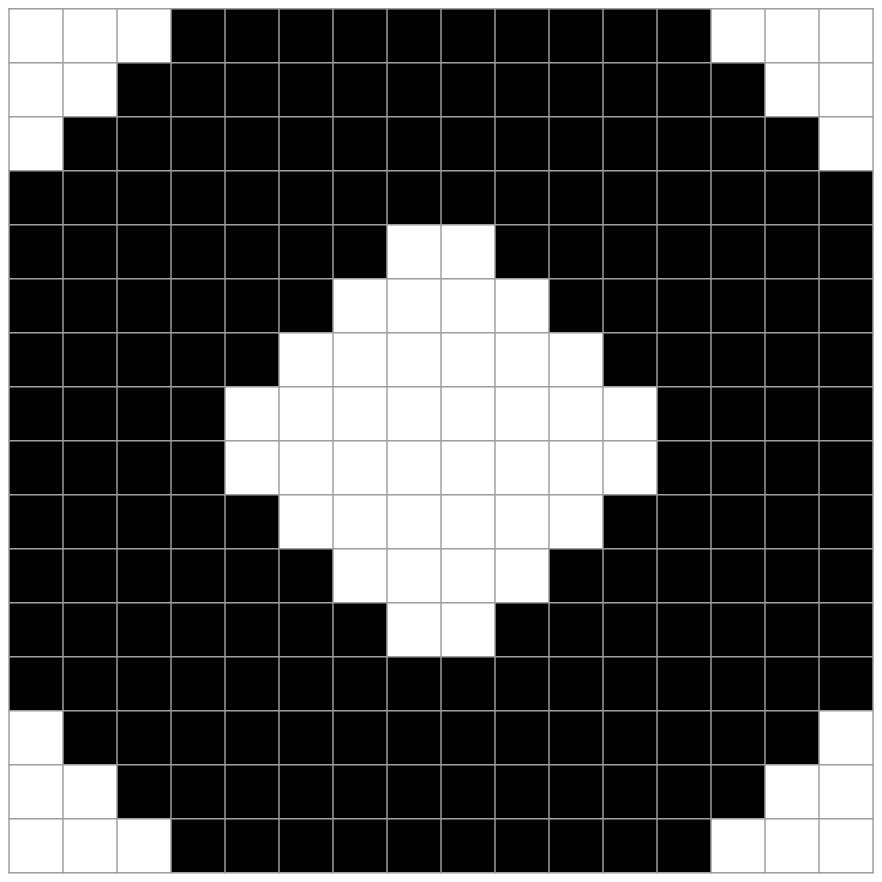}\\
   $n=11$
  \end{center}
  \label{fig:2dchaos-11}
 \end{minipage}
  \caption{Chaotic pattern. $W_n^{\vec{j}}$ with $p=q=1$.}\label{fig:2dchaos}
\end{figure}
\begin{figure}[htbp]
 \begin{minipage}{0.24\hsize}
  \begin{center}
   \includegraphics[width=29mm]{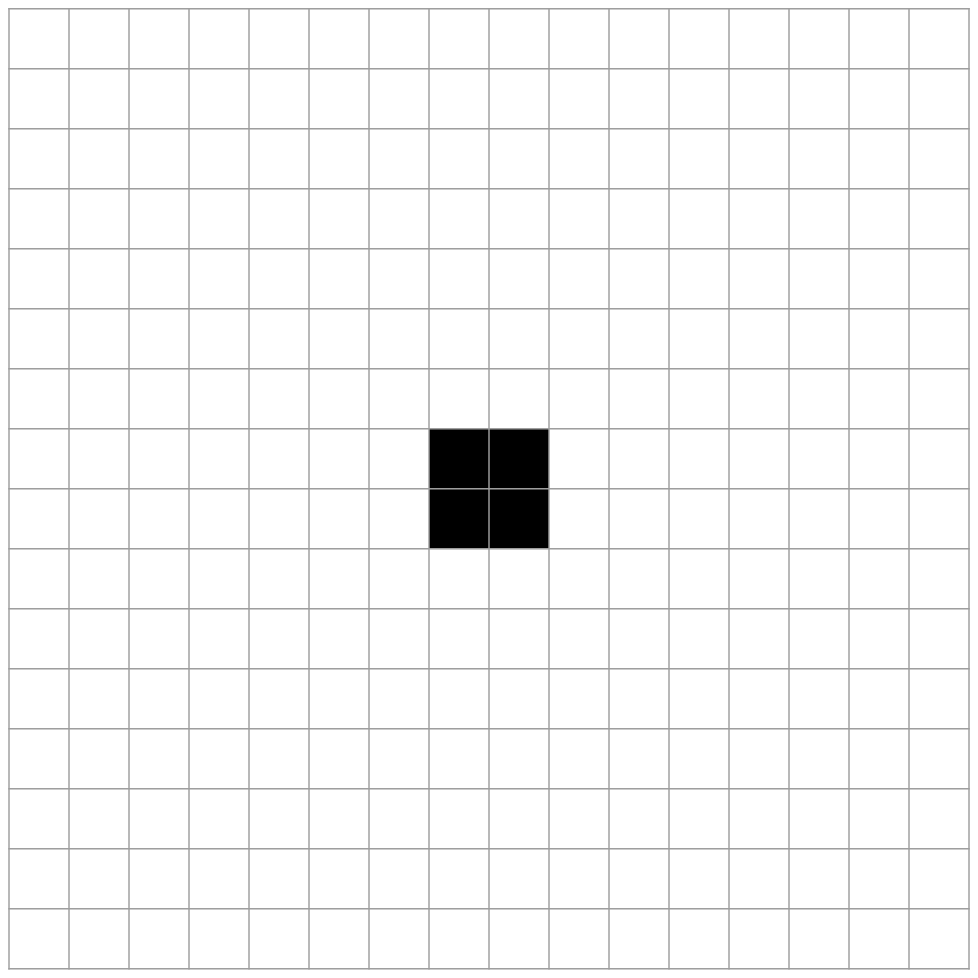}\\
   $n=0$
  \end{center}
  \label{fig:2by2-00}
 \end{minipage}
 \begin{minipage}{0.24\hsize}
  \begin{center}
   \includegraphics[width=29mm]{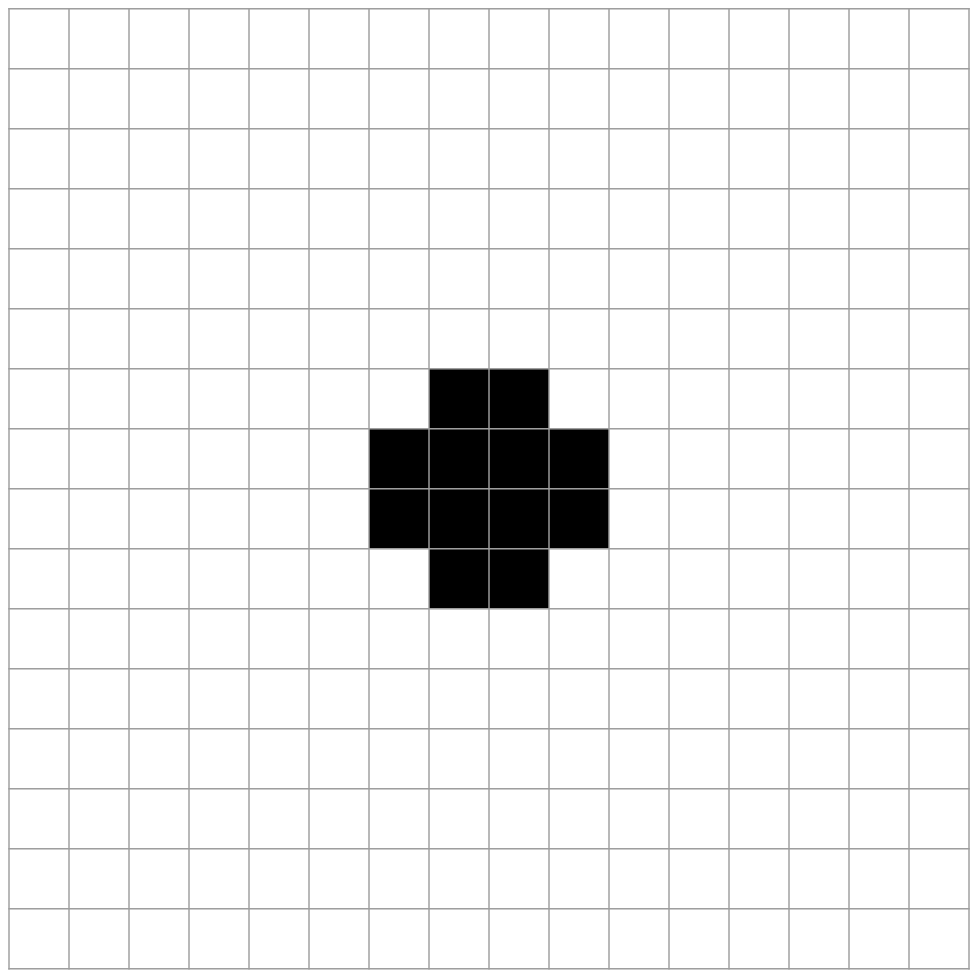}\\
   $n=1$
  \end{center}
  \label{fig:2by2-01}
 \end{minipage}
 \begin{minipage}{0.24\hsize}
  \begin{center}
   \includegraphics[width=29mm]{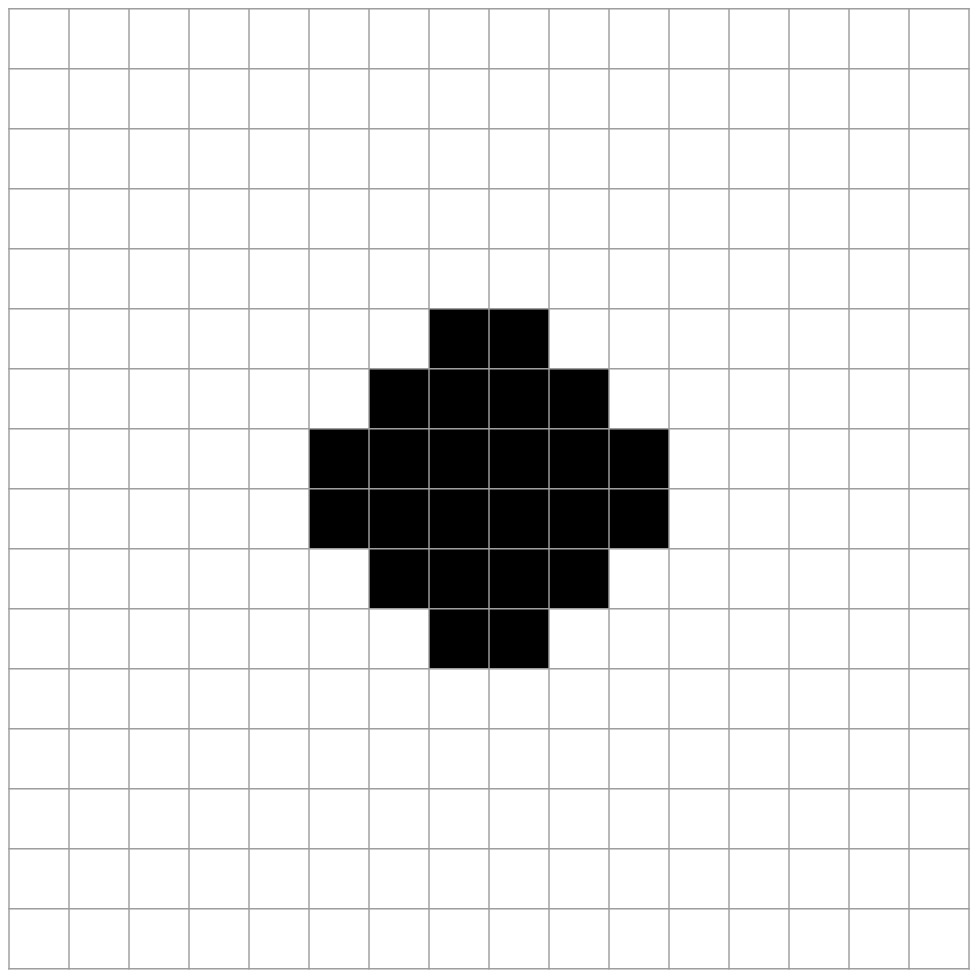}\\
   $n=2$
  \end{center}
  \label{fig:2by2-02}
 \end{minipage}
 \begin{minipage}{0.24\hsize}
  \begin{center}
   \includegraphics[width=29mm]{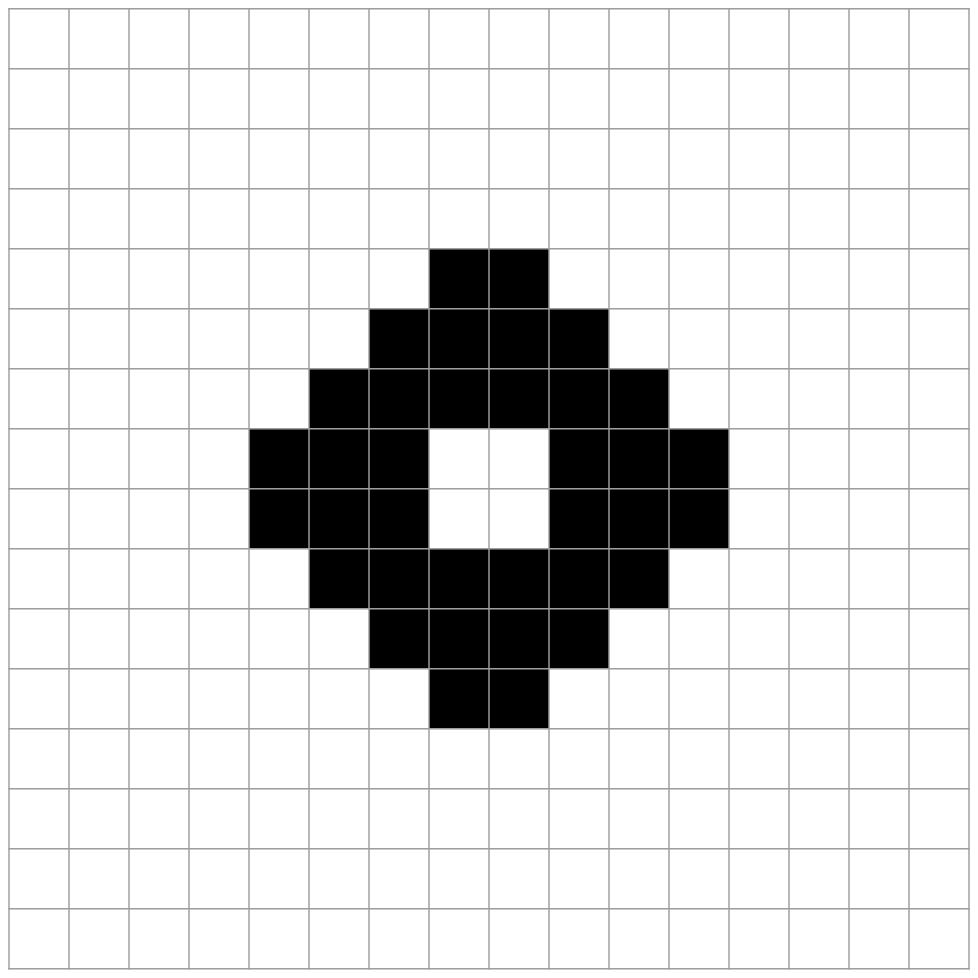}\\
   $n=3$
  \end{center}
  \label{fig:2by2-03}
 \end{minipage}\\
  \begin{minipage}{0.24\hsize}
  \begin{center}
   \includegraphics[width=29mm]{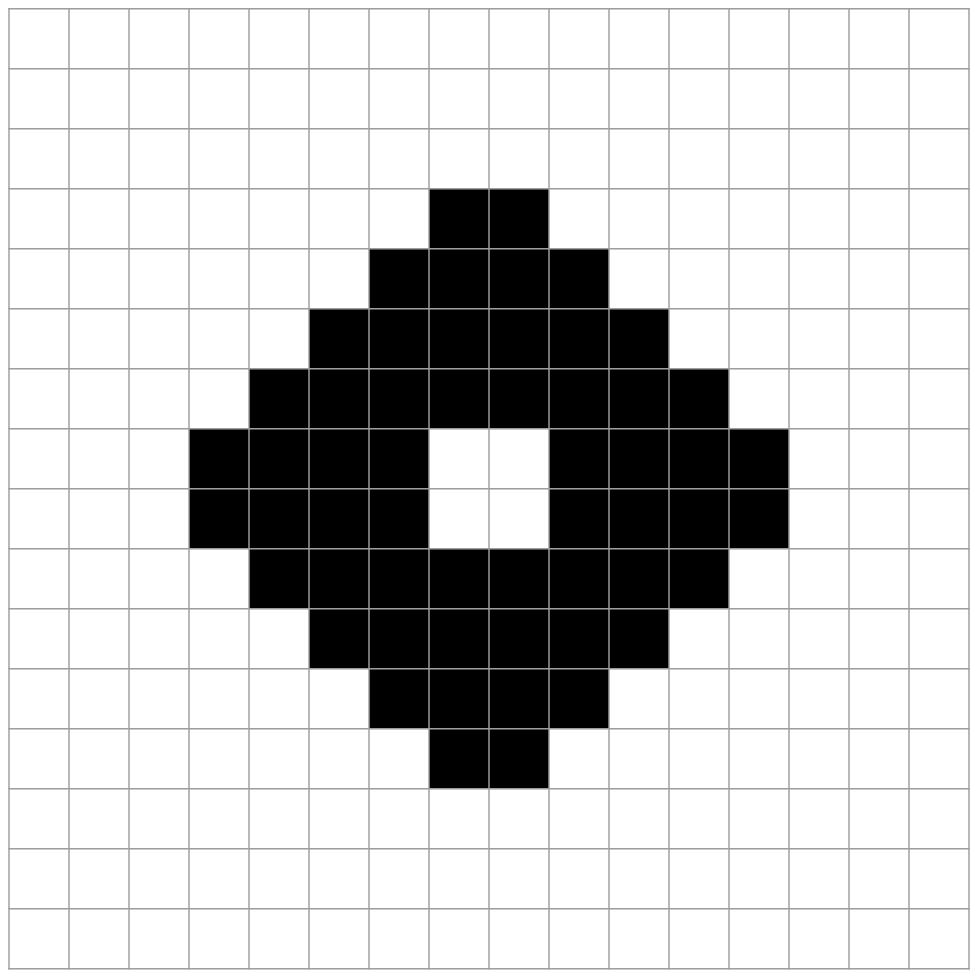}\\
   $n=4$
  \end{center}
  \label{fig:2by2-04}
 \end{minipage}
 \begin{minipage}{0.24\hsize}
  \begin{center}
   \includegraphics[width=29mm]{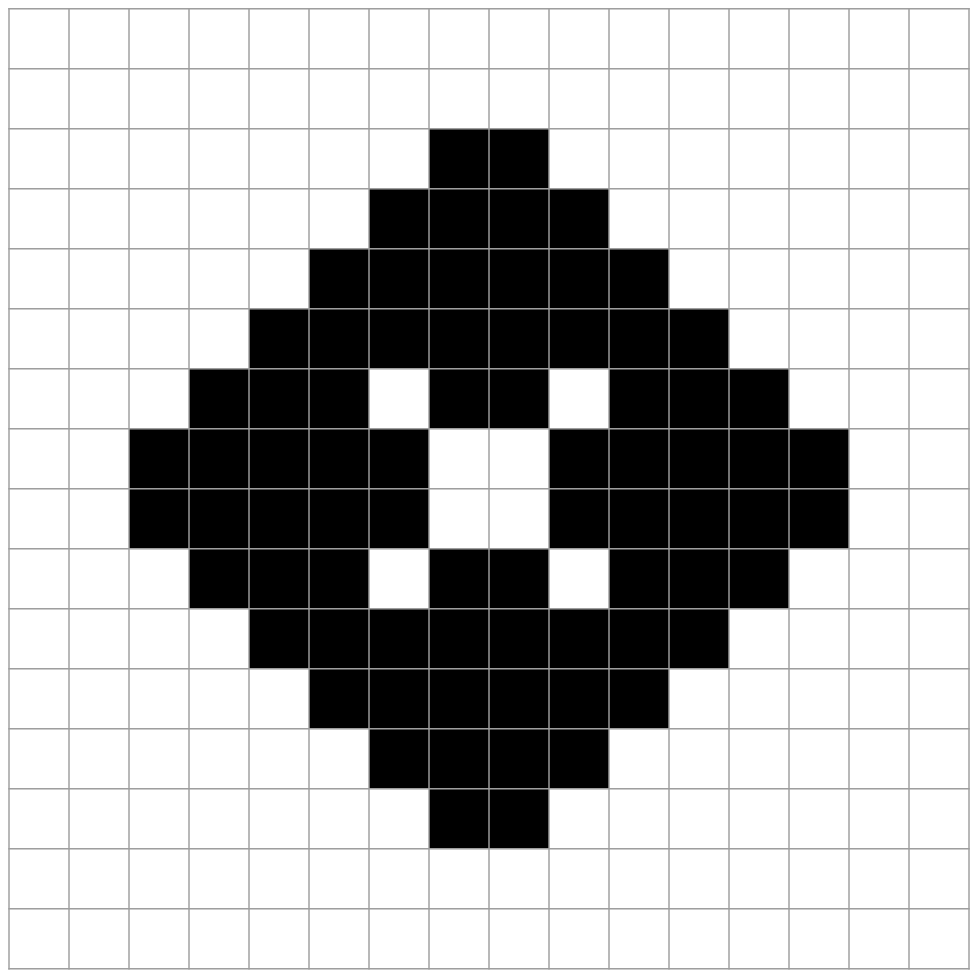}\\
   $n=5$
  \end{center}
  \label{fig:2by2-05}
 \end{minipage}
 \begin{minipage}{0.24\hsize}
  \begin{center}
   \includegraphics[width=29mm]{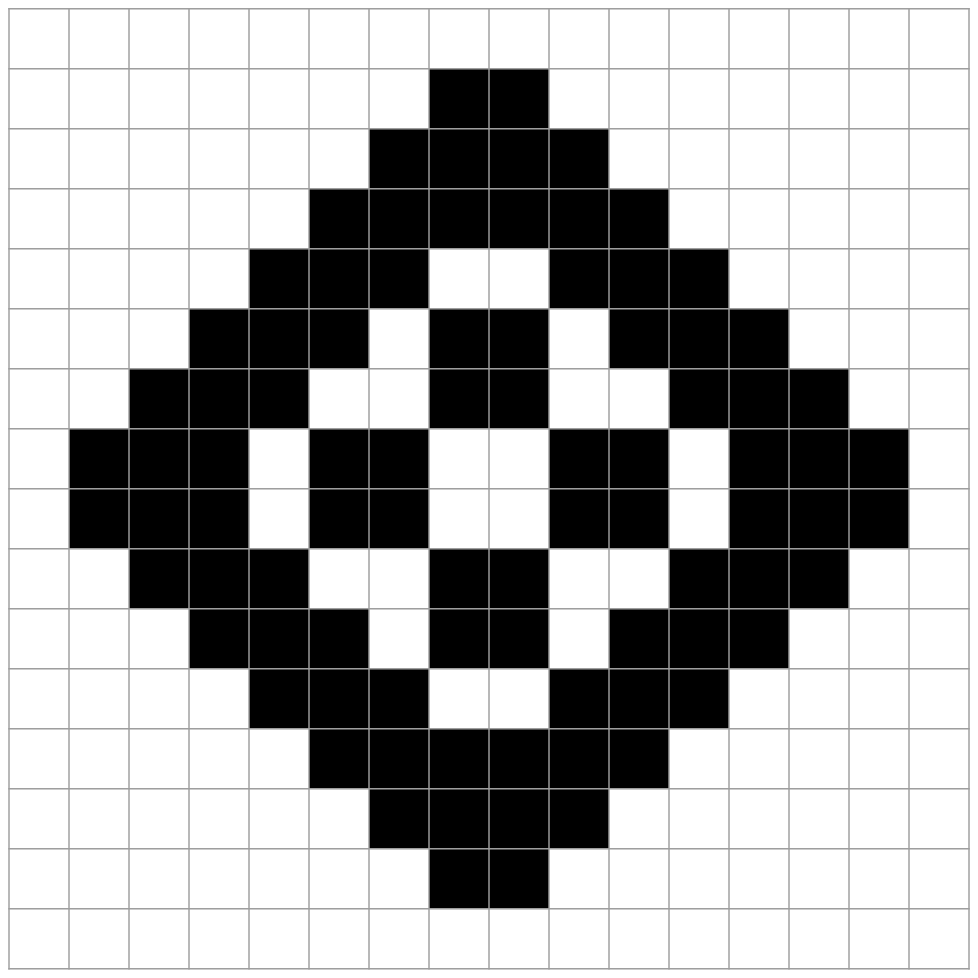}\\
   $n=6$
  \end{center}
  \label{fig:2by2-06}
 \end{minipage}
 \begin{minipage}{0.24\hsize}
  \begin{center}
   \includegraphics[width=29mm]{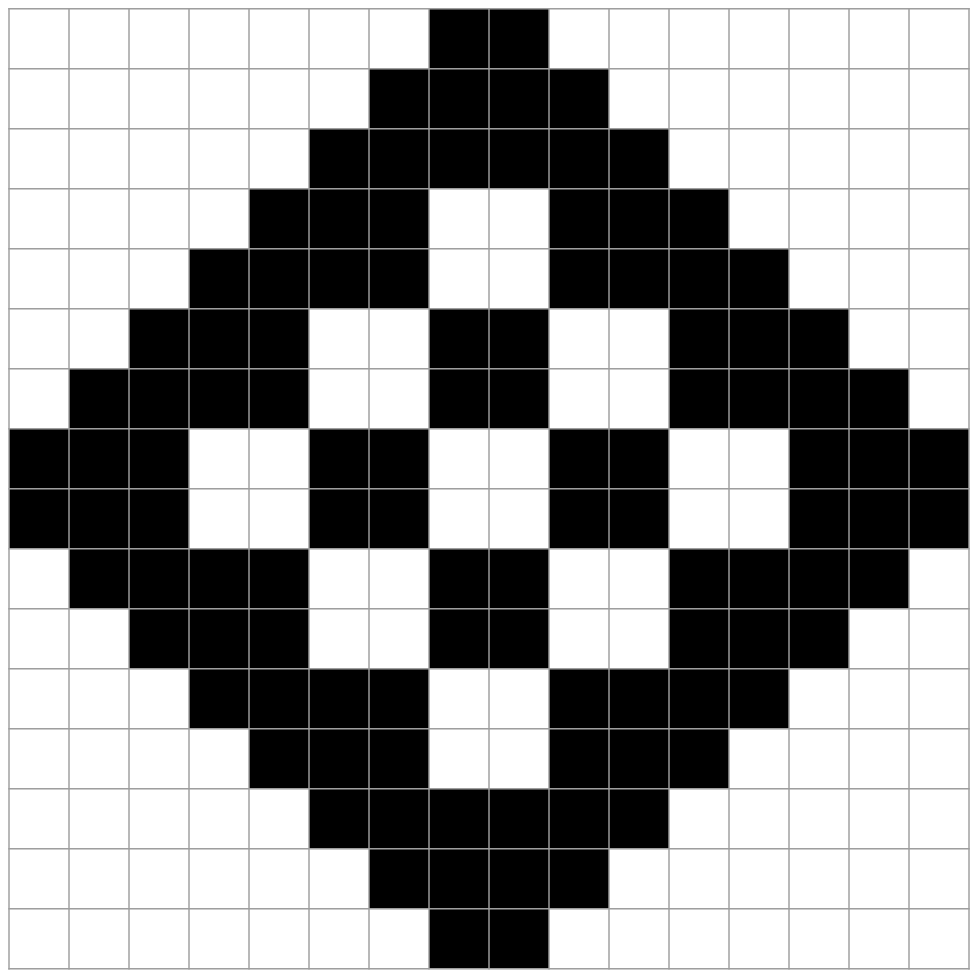}\\
   $n=7$
  \end{center}
  \label{fig:2by2-07}
 \end{minipage}\\
  \begin{minipage}{0.24\hsize}
  \begin{center}
   \includegraphics[width=29mm]{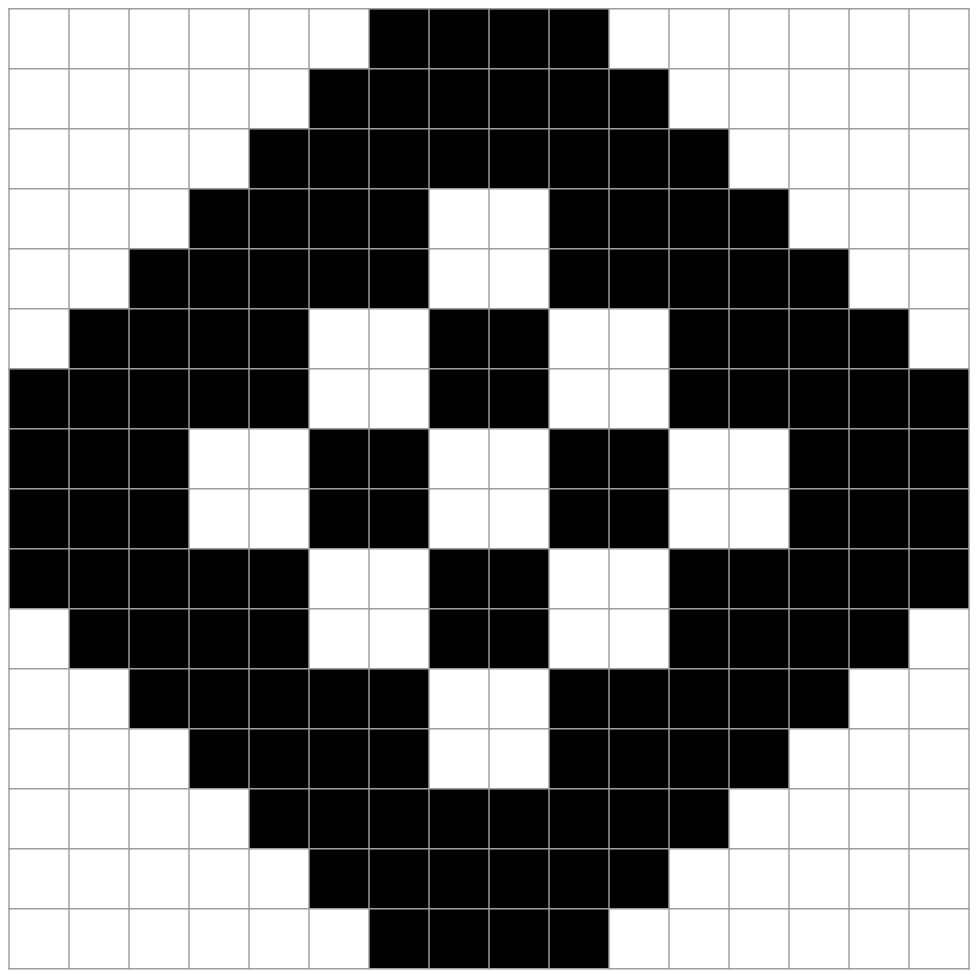}\\
   $n=8$
  \end{center}
  \label{fig:2by2-08}
 \end{minipage}
 \begin{minipage}{0.24\hsize}
  \begin{center}
   \includegraphics[width=29mm]{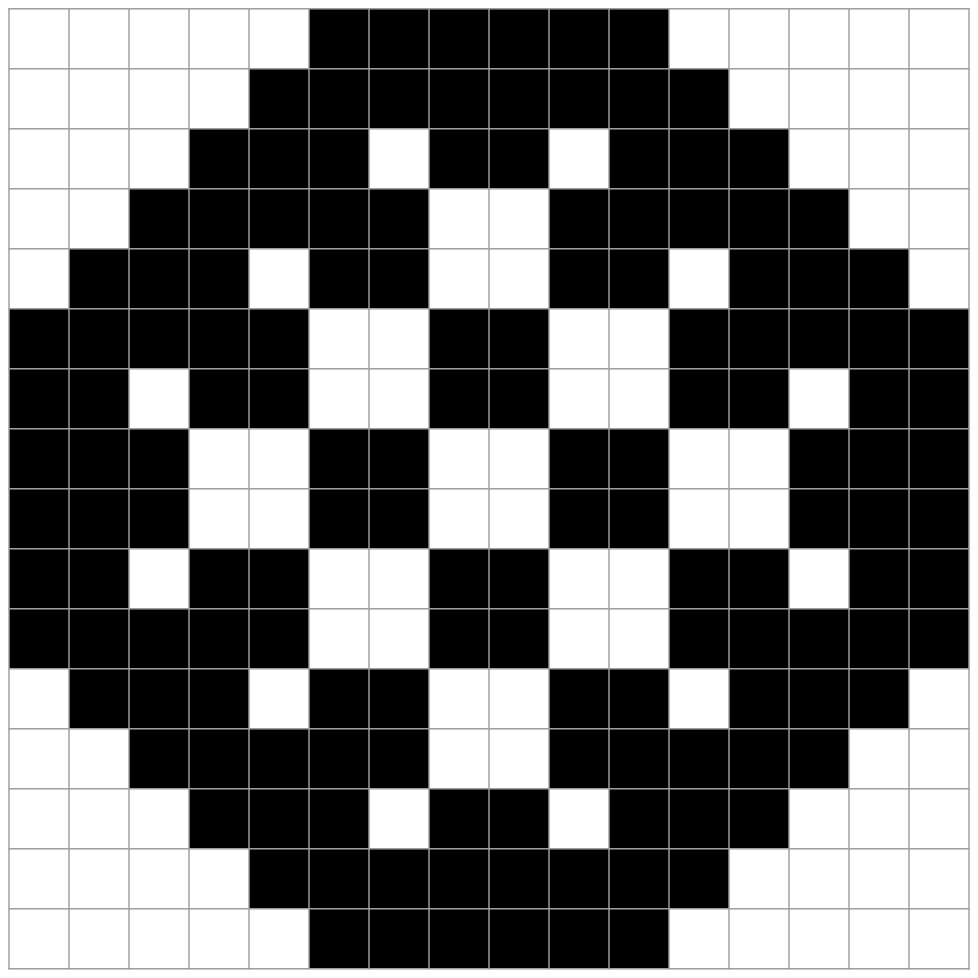}\\
   $n=9$
  \end{center}
  \label{fig:2by2-09}
 \end{minipage}
 \begin{minipage}{0.24\hsize}
  \begin{center}
   \includegraphics[width=29mm]{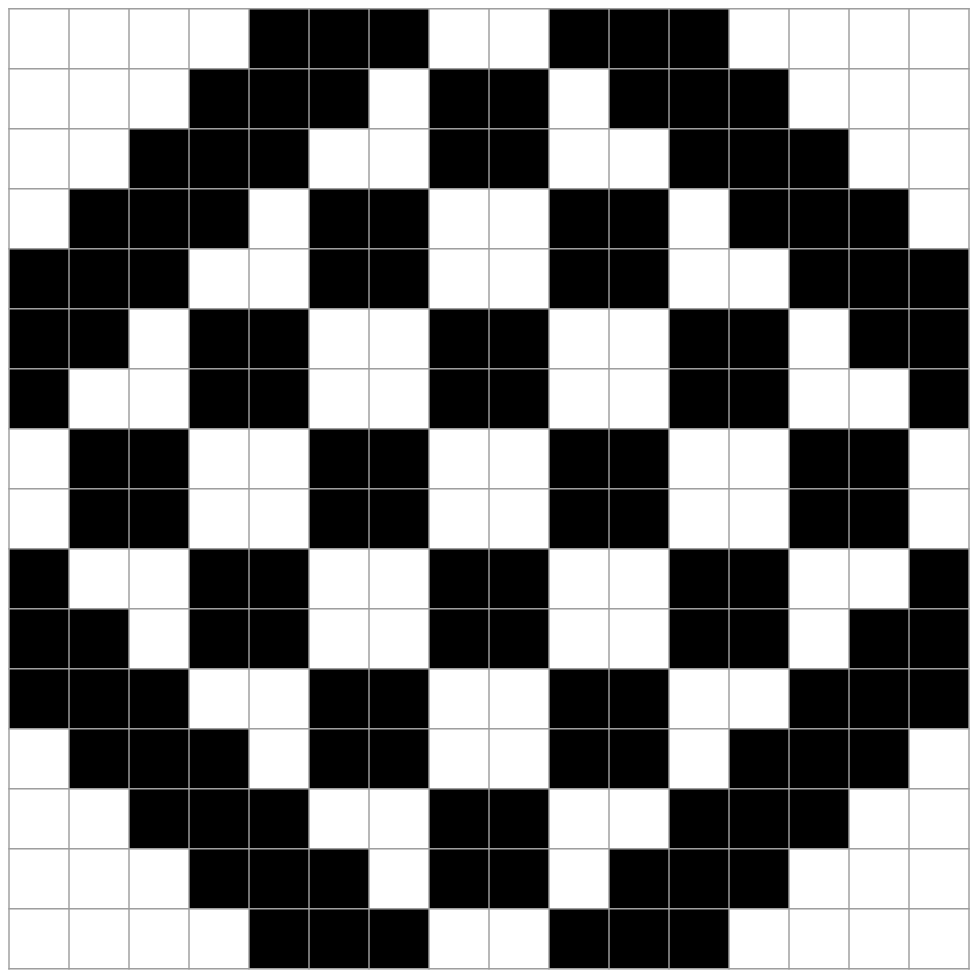}\\
   $n=10$
  \end{center}
  \label{fig:2by2-10}
 \end{minipage}
 \begin{minipage}{0.24\hsize}
  \begin{center}
   \includegraphics[width=29mm]{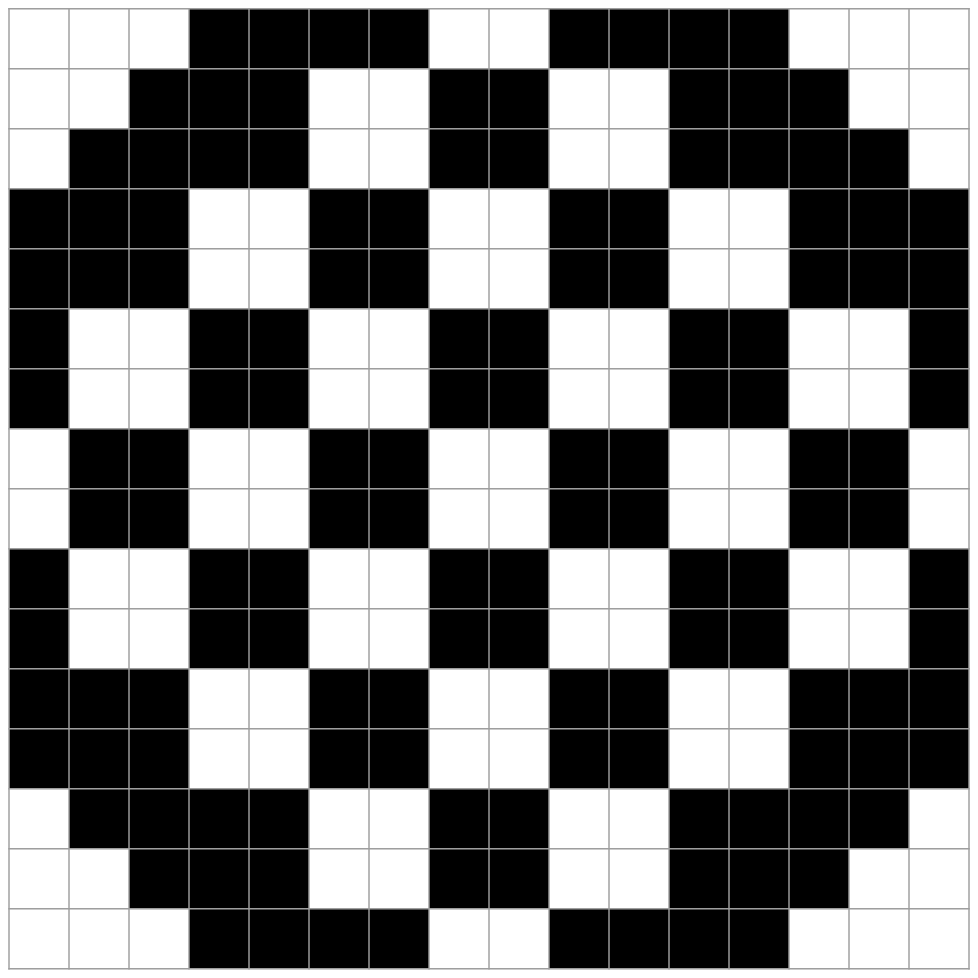}\\
   $n=11$
  \end{center}
  \label{fig:2by2-11}
 \end{minipage}
  \caption{Self-replicating pattern. $W_n^{\vec{j}}$ with $p=1$ and $q=2$.}\label{fig:2by2}
\end{figure}
\newpage
\section{Concluding remarks}
\label{s4}
In this article we proposed and investigated discrete and ultradiscrete Gray-Scott model, which is a two component reaction diffusion system.
We found that solutions of each equation reveal various spatial patterns.
Moreover, there are solutions of the discrete equation and the ultradiscrete equation which correspond to each other.
Indeed, the parameters $(a,b)$ with which the spatial pattern Figure \ref{3} is observed correspond to  the parameters $(A,B)$ with which the spatial pattern Figure \ref{fig:self} or Figure \ref{fig:self2}.
The ultradiscrete Gray-Scott model has a solution which is an elementary cellular automaton and which reveals Sierpinski gasket.
This is answer of the question ``What is the correspondence between cellular automata and continuous systems?" in \cite{10}.
Discrete equations and ultradiscrete equations whose solutions inherit properties of differential equations are studied in case of integrable equations.
We expect that more discretizations and ultradiscretizations which inherit properties of differential equations are studied and the various phenomena are made clear.
\section*{Acknowledgements}
The authors are deeply grateful to Prof. Tetsuji Tokihiro who provided helpful comments and suggestions.

This work was supported by JSPS KAKENHI Grant Number 23740125.

\end{document}